\pdfoutput=1
\RequirePackage{ifpdf}
\ifpdf 
\documentclass[pdftex]{sigma}
\else
\documentclass{sigma}
\fi

\newtheorem{Theorem}{Theorem}[section]

\newtheorem{Lemma}[Theorem]{Lemma}

{\theoremstyle{definition}
\newtheorem{Definition}[Theorem]{Definition}

\newtheorem{Example}[Theorem]{Example}
\newtheorem{Remark}[Theorem]{Remark}
}

\def\bu{{\bf u}}
\def\CC{{\mathbb C}}
\def\ZZ{{\mathbb Z}}

\numberwithin{equation}{section}

\begin{document}

\allowdisplaybreaks

\renewcommand{\thefootnote}{$\star$}

\renewcommand{\PaperNumber}{062}

\FirstPageHeading

\ShortArticleName{Af\/f\/ine and Finite Lie Algebras and Integrable Toda Field Equations}

\ArticleName{Af\/f\/ine and Finite Lie Algebras and Integrable Toda\\ Field Equations on Discrete Space-Time\footnote{This
paper is a contribution to the Special Issue ``Geometrical Methods in Mathematical Physics''. The full collection is available at \href{http://www.emis.de/journals/SIGMA/GMMP2012.html}{http://www.emis.de/journals/SIGMA/GMMP2012.html}}}

\Author{Rustem GARIFULLIN~$^\dag$, Ismagil HABIBULLIN~$^\dag$ and Marina YANGUBAEVA~$^\ddag$}

\AuthorNameForHeading{R.~Garifullin, I.~Habibullin and M.~Yangubaeva}

\Address{$^\dag$~Ufa Institute of Mathematics, Russian Academy of Science,\\
\hphantom{$^\dag$}~112~Chernyshevskii Str., Ufa, 450077, Russia}
\EmailD{\href{mailto:grustem@gmail.com}{grustem@gmail.com},
\href{mailto:habibullinismagil@gmail.com}{habibullinismagil@gmail.com}}

\Address{$^\ddag$~Faculty of Physics and Mathematics, Birsk State Social Pedagogical Academy,\\
\hphantom{$^\ddag$}~10~Internationalnaya Str., Birsk, 452452, Russia}
\EmailD{\href{mailto:marina.yangubaeva@mail.ru}{marina.yangubaeva@mail.ru}}

\ArticleDates{Received April 24, 2012, in f\/inal form September 14, 2012; Published online September 18, 2012}

\Abstract{Dif\/ference-dif\/ference systems are suggested corresponding to the Cartan matrices of any simple or af\/f\/ine Lie algebra. In the cases of the algebras $A_N$, $B_N$, $C_N$, $G_2$, $D_3$, $A_1^{(1)}$, $A_2^{(2)}$, $D^{(2)}_N$ these systems are proved to be integrable. For the systems corresponding to the algebras $A_2$, $A_1^{(1)}$, $A_2^{(2)}$ generalized symmetries are found. For the systems $A_2$, $B_2$, $C_2$, $G_2$, $D_3$ complete sets of independent integrals are found.  The Lax representation for the dif\/ference-dif\/ference systems corresponding to $A_N$, $B_N$, $C_N$, $A^{(1)}_1$, $D^{(2)}_N$ are presented.}


\Classification{35Q53; 37K40}


\renewcommand{\thefootnote}{\arabic{footnote}}
\setcounter{footnote}{0}

\section{Introduction}

Systems of partial dif\/ferential equations of the form
\begin{gather}\label{cont-system}
r_{x,y}^i=e^{\sum\limits_{j=1}^{j=N}{a_{ij}r^j}}, \qquad i=1,2,\dots,N,
\end{gather}
called generalized two-dimensional Toda lattices have very important applications in Liouville and conformal f\/ield theories, they are studied in details (see \cite{zamolodchikov,Bog,Cor,DrS,FG80,GanzhaTsarev,ZhiberGur,LeznovSavel'ev,LSSh1982,M,MOP,OP,ShabatYamilov} and the references therein). Here the matrix $A=\{a_{i,j}\}$ is the Cartan matrix of an arbitrary f\/inite or af\/f\/ine Lie algebra. It is known that in the former case system (\ref{cont-system}) is Darboux integrable while in the latter case~-- $S$-integrable. The widely known Drinfel'd--Sokolov formalism allows one to construct the Lax representation for the system (\ref{cont-system}) in terms of the Lie algebra canonically associated with the corresponding Cartan matrix~$A$.

The problem of f\/inding discrete versions of the system is intensively studied (see, for instance, \cite{AdlerStartsev,hab,H,KhG,HP,Sakieva,HZhS, Hirota,IH,KNS2,KNS,Suris,Tsuboi,Ward}). Recently, in \cite{HZY} integrable dif\/ferential-dif\/ference analog of system (\ref{cont-system}) was suggested
\begin{gather}\label{dis-cont-system}
v_{1,x}^i-v_{x}^i=e^{\sum\limits_{j=1}^{j=i-1}{a_{i,j}v^j}+\sum\limits_{j=i+1}^{j=N}{a_{i,j}v_1^j}+\frac12a_{i,i}(v^i+v^i_1)},\qquad i=1,2,\dots ,N.
\end{gather}
Here the functions $v^{j} = v^{j}(n,x)$, $ j=1, \dots , N$ are the searched f\/ield variables. The subindex denotes a shift of the discrete variable~$n$ or the derivative with respect to~$x$: $r_k^{j}=r^j(n+k,x)$ and $r_{x}^{j}=\frac{\partial}{\partial x}r^j(n,x)$. A particular case of (\ref{dis-cont-system}), corresponding to the algebra $C_N$ is also studied in~\cite{smirnov}.

In the present article we study the problem of further discretization of system~(\ref{cont-system}), i.e.\ the problem of f\/inding a rule allowing to assign to any Cartan matrix a system of integrable dif\/ference-dif\/ference equations approximating in the continuum limit system~(\ref{dis-cont-system}) and therefore system (\ref{cont-system}). As it was pointed out in \cite{Ward} the problem of  discretization is important from physical viewpoint, they might have applications in discrete f\/ield theory and in quantum physics (see also~\cite{IH,KNS,Suris}). They can also be regarded as dif\/ference schemes in numerical computations (see~\cite{KYN}).

Following questions were addressed in~\cite{Ward}:
\begin{enumerate}\itemsep=0pt
\item[1)] whether there exists an integrable discrete version  for any two dimensional Toda f\/ield equation (\ref{cont-system});

\item[2)] which kind of algebraic structure (like Lie algebra or Lie group) is naturally  related to discrete versions.
\end{enumerate}

The subject has intensively been studied during the last 10--15 years. Various discrete versions of Toda f\/ield equations were investigated in literature (see~\cite{H, IH,KNS2,KNS,Tsuboi}  and references therein). As an alternative answer to the f\/irst question we suggest a fully discrete version of system~(\ref{cont-system}) in the following form
\begin{gather}\label{dis-dis-system}
e^{-u_{1,1}^i+u_{1,0}^i+u_{0,1}^i-u_{0,0}^i}-1=e^{\sum\limits_{j=1}^{j=i-1}\!{a_{i,j}u_{0,1}^j}
+\sum\limits_{j=i+1}^{j=N}\!{a_{i,j}u_{1,0}^j}+\frac12a_{i,i}(u_{0,1}^i+u_{1,0}^i)},\!\qquad i=1,2,\dots ,N,\!\!
\end{gather}
which evidently approximates (\ref{cont-system}) and (\ref{dis-cont-system}). Here $u^j=u^j(n,m)$, $j=1,2,\dots, N$, is a set of the f\/ield  variables. The subindex indicates shifts of the arguments~$n$,~$m$ as follows $u^j_{i,k}:=u^j(n+i,m+k)$. The system corresponding to the algebra~$A_N$ coincides with that found years ego by Hirota (see~\cite{Hirota}). Obviously, system~\eqref{dis-dis-system} is invariant under the replacement~$n\leftrightarrow -m$. This property is inherited from the fact that system \eqref{cont-system} is invariant under the change $x\leftrightarrow y$.

The main result of the present article is in formulating of the conjecture below and proving it for numerous examples of Lie algebras.

\medskip

\noindent
{\bf Conjecture.}
\begin{enumerate}\itemsep=0pt\it
\item[$a)$] If $A$ is the Cartan matrix of a semi-simple Lie algebra then \eqref{dis-dis-system} is  Darboux integrable, in other words, it admits a complete set of integrals in both directions. Roughly speaking Darboux integrable difference-difference equations can be  reduced to ordinary difference equations.
\item[$b)$] If $A$ is the Cartan matrix of an affine Lie algebra then \eqref{dis-dis-system} is $S$-integrable, i.e.\ it can be integrated by means of the inverse scattering transform method. It is generally accepted that existence of generalized symmetries or Lax pairs indicates $S$-integrability.
\end{enumerate}

Part $a)$ of the conjecture is proved for the cases: $A_2$, $B_2$, $C_2$, $G_2$, $D_3$ by f\/inding complete sets of integrals. Systems corresponding to the algebras $A_N$, $B_N$, $C_N$ are studied by using Lax pair in Section~\ref{section6}. In the cases $A_2$, $B_2$ ($C_2$), $G_2$ the exponential system (\ref{dis-dis-system}) reads as follows
\begin{gather*}
e^{-u_{1,1}^1+u_{1,0}^1+u_{0,1}^1-u_{0,0}^1}-1=e^{u^1_{0,1}+u_{1,0}^1-u_{1,0}^2},\qquad
e^{-u_{1,1}^2+u_{1,0}^2+u_{0,1}^2-u_{0,0}^2}-1=e^{-cu^1_{0,1}+u_{0,1}^2+u_{1,0}^2},
\end{gather*}
where $c$ takes only three values $c=1, 2, 3$. Here the values $c=1$, $c=2$, $c=3$ correspond to the algebras $A_2$, $B_2$ or $C_2$, $G_2$ respectively.
Let us introduce a notation for the nonlinear second-order partial dif\/ference operator by setting
\begin{gather*}
\Delta(u):=e^{-u_{1,1}+u_{1,0}+u_{0,1}-u_{0,0}}-1,
\end{gather*}
then the last system takes a compact form
\begin{gather}\label{Delta-sys}
\Delta\big(u^1\big)=e^{u^1_{0,1}+u_{1,0}^1-u_{1,0}^2}, \qquad
\Delta\big(u^2\big)=e^{-cu^1_{0,1}+u_{0,1}^2+u_{1,0}^2}.
\end{gather}
Integrals in both directions for system (\ref{Delta-sys}) are given in Section~\ref{section2}.

Since the map def\/ined in (\ref{dis-dis-system}) converts any $N\times N$ matrix to a system of dif\/ference-dif\/ference equations one can easily specify the form of systems corresponding to any Cartan matrix cano\-ni\-cally related to a f\/inite or af\/f\/ine Lie algebra both of classical series or exceptional. For instance, for the algebra $D_3$ with the Cartan matrix
\begin{gather*}
A=
\begin{pmatrix}
2&-1&-1\\
-1&2&0\\
-1&0&2
\end{pmatrix},
\end{gather*}
we have an integrable system of the form
\begin{gather*}
\Delta\big(u^1\big)=e^{u^1_{0,1}+u_{1,0}^1-u_{1,0}^2-u_{1,0}^3}, \qquad
\Delta\big(u^2\big)=e^{-u^1_{0,1}+u_{0,1}^2+u_{1,0}^2},\qquad
\Delta\big(u^3\big)=e^{-u^1_{0,1}+u_{0,1}^3+u_{1,0}^3}.
\end{gather*}
Complete sets of integrals for this system is given in Section~\ref{section2}.

For the Kac--Moody algebra $A^{(2)}_2$ with the Cartan matrix
\begin{gather*}
A=
\begin{pmatrix}
2&-1\\
-4&2
\end{pmatrix}
\end{gather*}
the corresponding system is
\begin{gather}\label{Delta-sys2}
\Delta\big(u^1\big)=e^{u^1_{0,1}+u_{1,0}^1-u_{1,0}^2},\qquad
\Delta\big(u^2\big)=e^{-4u^1_{0,1}+u_{0,1}^2+u_{1,0}^2}.
\end{gather}

For the dif\/ference-dif\/ference systems corresponding to the algebras  $A_2$, $A_1^{(1)}$, $A_2^{(2)}$ genera\-li\-zed symmetries are found, for $A_2$ the symmetries have usual form, for $A_1^{(1)}$, $A_2^{(2)}$ they have nonlocal form (or, they are of hyperbolic type, see Def\/inition~\ref{sh} below). In the literature generalized symmetries are regarded as a criterion of integrability. Thus one concludes that sys\-tems~(\ref{Delta-sys2}),~(\ref{disa22}) provide new examples of $2\times 2$ $S$-integrable quad graph models.

Stress that the number of the f\/ield variables in all three systems  (\ref{cont-system}), (\ref{dis-cont-system}), (\ref{dis-dis-system}) coincides with the rank of the corresponding Lie algebra.

It is known that any system of dif\/ference-dif\/ference equations of hyperbolic type admits a~pair of characteristic Lie algebras which are ef\/fectively evaluated (see \cite{hab}). Characteristic Lie algebras of system (\ref{dis-dis-system}) are closely connected with the Lie algebra canonically related to the given Cartan matrix~$A$. In our opinion this relation could allow one to answer the second question in~\cite{Ward} (see the list of questions above).

The problem of developing discrete versions of the Drinfel'd--Sokolov formalism is also challenging since today systems of discrete equations are very popular. They  have a large variety of application in theoretical physics, in discrete geometry, in the theory of
tau functions in lattice Toda f\/ield equations. The present article would also provide an ``experimental'' background for creating a discrete theory parallel to the Drinfel'd--Sokolov formalism.

The article is organized as follows. In Section~\ref{section1.1} notions of the integrals and generalized symmetries of both evolutionary and hyperbolic type for quad graph equations are def\/ined. Suf\/f\/icient condition of complete set of integrals is proved. In Section~\ref{section2} a method of discretization of Darboux integrable models preserving integrability is discussed. The method is explained in details by an example of the Liouville equation \cite{Sakieva}. Then it is applied to the system~(\ref{dis-cont-system}) to get its dif\/ference-dif\/ference analog. The case~$B_2$ is studied in more details. These simulations allowed  us to guess the formula~(\ref{dis-dis-system}). Complete sets of independent integrals for the systems $A_2$, $B_2$, $C_2$, $G_2$, $D_3$ are presented.

Generalized symmetries are evaluated for the systems corresponding $A_2$, $A_1^{(1)}$, $A_2^{(2)}$ in Section~\ref{section3}. Special attention is paid to hyperbolic type symmetries which provide a semi-discrete version of the consistency around a cube property of quad graph equations~\cite{ABS}. In Section~\ref{section4} the concept of the characteristic Lie algebra (see~\cite{hab}) of discrete models is brief\/ly discussed. Here a~complete description of such algebra is obtained for the system of dif\/ference-dif\/ference equations~$A_2$.

In Section~\ref{section5} cutting of\/f conditions for the Hirota equation preserving integrability are studied by using the method suggested in \cite{H}. It turned out that some of the systems in the class (\ref{dis-dis-system}) can be obtained from the Hirota equation by imposing proper boundary conditions. Since the cutting of\/f conditions are compatible with the Lax representation one can derive together with the reduced discrete system also its Lax pair. In Section~\ref{section5} the Lax pairs for the dif\/ference-dif\/ference systems $A_N$, $B_N$, $C_N$, $A_1^{(1)}$, $D_N^{(2)}$ are found.

In Section~\ref{section6} an algorithm is suggested to look for integrals via the Lax pair. In Section~\ref{section7} widely known periodical reduction of the Hirota equation is discussed. The Lax pair for this system found in \cite{Zabr} is rewritten in terms of the Cartan--Weyl basis for the algebra $A_N^{(1)}$. Emphasize that the dif\/ference-dif\/ference system obtained as a periodical reduction and that given by the formula (\ref{dis-dis-system}) corresponding to $A_N^{(1)}$ are not equivalent (see Remark~\ref{period} below).

\subsection{Integrals and symmetries for quad graph systems}\label{section1.1}

Consider a system of quad graph equations of general form
\begin{gather}\label{sg}
H({\bf u}_{n,m},{\bf u}_{n+1,m},{\bf u}_{n,m+1},{\bf u}_{n+1,m+1})=0,
\end{gather}
where $\bu=\bu_{n,m}$ is a vector-function depending on two integers and ranging on $\CC^N$: $\bu=(u^1,u^2,\ldots, u^{N})^T.$ As usually we request that equation~\eqref{sg} can be solved with respect to any of the arguments $\bu_{n,m}$, $\bu_{n+1,m}$, $\bu_{n,m+1}$, ${\bf u}_{n+1,m+1}$. In other words there exists a set of functions $H^{(\pm 1,\pm 1)}$ such that
\begin{gather*}
\bu_{n+1,m+1}=H^{1,1}({\bf u}_{n,m},{\bf u}_{n+1,m},{\bf u}_{n,m+1}),\\
\bu_{n-1,m+1}=H^{(-1,1)}(\bu_{n-1,m},\bu_{n,m},\bu_{n,m+1}),\\
\bu_{n+1,m-1}=H^{(1,-1)}(\bu_{n+1,m},\bu_{n,m},\bu_{n,m-1}),\\
\bu_{n-1,m-1}=H^{(-1,-1)}(\bu_{n-1,m},\bu_{n,m},\bu_{n,m-1}).
\end{gather*}

Def\/ine the following standard set of dynamical variables which consists of the variable $\bu_{n,m}$ and its shifts $\bu_{n+k,m}$ and $\bu_{n,m+l}$ where $k,l\in \ZZ$
\[
S=\{\bu_{n+k,m},\bu_{n,m+l}: \, k,l\in \ZZ\} .
\]
By $[\bu]$ we denote a f\/inite set of dynamical variables, for instance notation $h=h([\bu])$ means that function $h$ depends on a f\/inite number of dynamical variables.
Def\/ine shift operators $D_m$, $D_n$ acting due to the rules
\[
D_mf(n,m)=f(n,m+1),\qquad D_nf(n,m)=f(n+1,m).
\]

\begin{Definition}\null \qquad
\begin{enumerate}\itemsep=0pt
\item[$i)$] Function $F([\bu],n,m)$ (function $I([\bu],n,m)$) is called $m$-integral (respectively $n$-integral) of the equation \eqref{sg} if the following identity holds $D_m F=F$ (or, $D_n I=I$) on arbitrary solutions of equation~\eqref{sg}.
\item[$ii)$] Integrals of the form $F=F(n)$ ($I=I(m)$) are called trivial.
\item[$iii)$] Equation admitting  $N$  non-trivial independent integrals in each direction is called Darboux integrable.
\item[$iv)$] Set of integrals is called independent if none of them can  be expressed through the other integrals and their shifts.
\end{enumerate}
\end{Definition}

It can be easily proved that $m$-integrals do not depend on the variables $u_{n,m+l}$ with $l\neq 0$ and similarly $n$-integrals do not depend on $u_{n+k,m}$ where $k\neq 0$.

Now formulate a very simple and convenient suf\/f\/icient condition of complete set of integrals.

\begin{Theorem}\label{independent_int}
Let us given a set of $m$-integrals of the form
\begin{gather}\label{Ij}
I_{(j)}=I_{(j)}\big(n, m, {\bf u},D_{n}{\bf u},D_{n}^2{\bf u},\ldots,D_{n}^{\gamma_j}{\bf u}\big),\qquad \gamma_j\geq 0, \qquad j=1,\ldots,N.
\end{gather}
Suppose that for any $j$ and for all $n$, $m$ at least one of the derivatives $\frac{\partial I_{(j)}}{\partial u^i}$ differs from zero and the condition
\begin{gather}\label{J_det}
\det\left(\frac{\partial I_{(j)}}{\partial u^i}\right)\neq 0
\end{gather}
holds for all $n$, $m$. Then the integrals constitute a complete set of integrals.
\end{Theorem}

\begin{proof}
Suppose in contrary that the set of integrals (\ref{Ij}) is not independent. Then at least one of the integrals, say for def\/initeness $I_{(1)}$ is a function of the form
\begin{gather}\label{not_ind}
I_{(1)}=Q\big(I_{(2)},I_{(3)},\ldots,I_{(N)}, D_nI_{(2)},D_nI_{(3)},\ldots,D_nI_{(N)},\ldots\big),
\end{gather}
depending on the other integrals and their shifts.
Dif\/ferentiating (\ref{not_ind}) with respect to $u^i$ we obtain the following equalities
\begin{gather*}
\frac{\partial I_{(1)}}{\partial u^i}-\frac{\partial I_{(2)}}{\partial u^i}\frac{\partial Q}{\partial I_{(2)}}-
\cdots -\frac{\partial I_{(N)}}{\partial u^i}\frac{\partial Q}{\partial I_{(N)}}=0.
\end{gather*}

The latter can be regarded as a linear algebraic system with the coef\/f\/icient matrix $\{\frac{\partial I_{(j)}}{\partial u^i}\}$ and a solution $\left(1,\frac{\partial Q}{\partial I_{(2)}},\ldots,\frac{\partial Q}{\partial I_{(N)}}\right)$ which obviously is not trivial. Therefore according to the well-known theorem the determinant of the coef\/f\/icient matrix should vanish. But it contradicts~(\ref{J_det}). Thus our assumption that the set of integrals is dependent is not correct. Theorem is proved.
\end{proof}

\begin{Remark}
Condition (\ref{J_det}) actually means that jacobian of the map ${\bf u}\rightarrow {\bf I}=\left(I_{(1)},\ldots,I_{(N)}\right)$ dif\/fers from zero for any value of the variables $\bf u, D_{n}{\bf u},D_{n}^2{\bf u},\ldots,D_{n}^{\gamma}{\bf u}$ ranging in a domain. Here $\gamma=\max\{\gamma^j\}$. Note that~(\ref{J_det}) is not a necessary condition for independent integrals. In the Example~\ref{counterex} we have a pair of independent integrals $I_{(1)}$ and $I_{(2)}$ for which condition~(\ref{J_det}) is violated.
\end{Remark}

\begin{Definition}\label{se}
An equation of the form
\begin{gather*}
\frac{d}{dt}u_{n,m}=G([\bu])
\end{gather*}
is called generalized symmetry of \eqref{sg} if the following compatibility condition is satisf\/ied
\begin{gather*}\frac{d}{dt}\big(u_{n+1,m+1}-F^{(1,1)}\big)\big|_{F=0,u_{n,m,t}=G}=0.
\end{gather*}
\end{Definition}

It can be shown that function $G([\bu])$ is a solution of linearized equation
\begin{gather*}
\frac{\partial F}{\partial u_{n+1,m+1}}D_nD_m G+\frac{\partial F}{\partial u_{n+1,m}}D_n G+\frac{\partial F}{\partial u_{n,m+1}}D_m G+\frac{\partial F}{\partial u_{n,m}}G=0.
\end{gather*}

There are systems of the form \eqref{sg} which do not admit evolutionary type symmetries, however they admit symmetries of more complicated structure. We call them hyperbolic type symmetries. Such type symmetries for~\eqref{sg} consist of two semi-discrete equations
\begin{gather}
\label{hs}
\frac{d}{dt}\bu_{n+1,m}=G\left(\frac{d}{dt}\bu_{n,m},[\bu]\right),\qquad
 \frac{d}{dt}\bu_{n,m+1}=\widetilde G\left(\frac{d}{dt}\bu_{n,m},[\bu]\right).
\end{gather}

\begin{Definition}\label{sh} Pair of equations \eqref{hs}  def\/ine a hyperbolic type symmetry for \eqref{sg} if equations~(\ref{sg}),~(\ref{hs}) constitute a commuting triple, i.e.\ the following compatibility conditions are satisf\/ied
\[
\frac{d}{dt}F^{(1,1)}=D_m G=D_n \widetilde G
\]
 by means of equations (\ref{sg}), (\ref{hs}).
\end{Definition}

Hyperbolic type symmetries are constructed in Section~\ref{section3}. They are nothing else but non-local symmetries (see, for instance,~\cite{Sh95}) rewritten in a  more convenient form. In the other hand side existence of hyperbolic type symmetries \eqref{hs} for equation \eqref{sg} can be regarded as a semi-discrete generalization of the well-known property of consistency around a cube~\cite{ABS, BobenkoSuris,Nijhoff} which is approved to be a criterion of integrability.

\section{Systematic approach to the problem of discretization\\ of the Darboux integrable systems}\label{section2}

The problem of f\/inding integrable discretizations of the integrable partial dif\/ferential equation is very complicated and not enough studied. The same is true for evaluating the continuum
limit for discrete models. In \cite{Sakieva} an ef\/fective algorithm of discretization (as well as evaluation of the continuum limit) of Darboux integrable equations is suggested based on the integrals. In this section we discuss the essence of the algorithm and apply it to exponential type systems. Note that sets of integrals for the systems corresponding to $A_2$, $B_2$, $G_2$, $D_3$ considered in this section as easily proved by applying suf\/f\/icient condition given in Theorem~\ref{independent_int} are independent.

\subsection{Explanation of the method with example of Liouville equation}

Its well-known that the famous Liuoville equation $u_{xy}=e^u$
admits integrals in both directions: $W=u_{xx}-0.5u_x^2$ and  $\bar{W}=u_{yy}-0.5u_y^2$. Indeed, it is very easy to check that $D_yW=0$ and $D_x\bar W=0$.

Consider now the problem of f\/inding all chains of the form
\begin{gather}\label{semi-discrete}
t_x(n+1,x)=f(x,t(n,x),t(n+1,x),t_x(n,x)) ,
\end{gather}
having the function $I=t_{xx}-\frac{1}{2}{t_x}^2$ as their $n$-integral. Equality $D_nI=I$ implies
\begin{gather}\label{(1)}
f_x+f_tt_x+f_{t_1}f+f_{t_x}t_{xx}-\frac{1}{2}f^2=t_{xx}-\frac{1}{2}{t_x}^2.
\end{gather}
By comparing the coef\/f\/icients before $t_{xx}$ in (\ref{(1)}) we have
$f_{t_x}=1$. Therefore,
\begin{gather}\label{(2)}
f(x,t,t_1,t_x)=t_x+d(x,t,t_1) .
\end{gather}
We substitute (\ref{(2)}) into (\ref{(1)}) and get
$d_x+d_tt_x+d_{t_1}t_x+d_{t_1}d-\frac{1}{2}{t_x}^2-dt_x-\frac{1}{2}d^2=-\frac{1}{2}{t_x}^2$,
or equivalently,
$d_t+d_{t_1}-d=0$
  and $d_x+d_{t_1}d-\frac{1}{2}d^2=0$.
   We solve the last two equations simultaneously and f\/ind that   $d=Ce^{\frac{1}{2}(t_1+t)}$ and $C$ is an arbitrary
   constant. Therefore, chain~(\ref{semi-discrete}) with
   $n$-integral $I=t_{xx}-\frac{1}{2}{t_x}^2$ becomes
   $t_{1x}=t_x+Ce^{(t_1+t)/2}$. The last equation in its turn admits also an $x$-integral $F=e^{(t_1-t)/2}+e^{(t_1-t_2)/2}$.

The next step consists in describing equations of the form
\begin{gather}\label{discrete}
v(n+1,m+1)=f(v(n,m),v(n+1,m),v(n,m+1))
\end{gather}
with $m$-integral  $F=e^{v_1-v}+e^{v_1-v_2}$.
Denote  $w_{ij}:=e^{-v_{ij}}$. In the new variables
$F=\frac{w_{0,0}+w_{2,0}}{w_{1,0}}$ is an $m$-integral of
equation $w_{1,1}=g(w_{0,0},w_{1,0},w_{0,1})$.
 $D_m F=F$ implies
 \begin{gather}\label{(28)}
\frac{w_{2,0}+w_{0,0}}{w_{1,0}}=\frac{D_ng+w_{0,1}}{g}.
 \end{gather}
 We dif\/ferentiate both sides of (\ref{(28)}) with respect to $w_{2,0}$ and apply the shift operator $D^{-1}_n$, we have
 \[
\frac{1}{w_{1,0}}=\frac{\partial_{{w_{2,0}}}{D_n g}}{g} \quad \Rightarrow \quad
 D^{-1}_n\left(\frac{1}{w_{1,0}}\right)=D^{-1}_n\left(\frac{\partial_{{w_{2,0}}}{D_n g}}{g}\right)\quad \Rightarrow \quad
g_{w_{1,0}}=\frac{w_{0,1}}{w_{0,0}}.
\]
Therefore,
 \begin{gather}\label{(30)}
  g=\frac{w_{0,1}w_{1,0}}{w_{0,0}}+c(w_{0,0},w_{0,1}).
\end{gather}
We substitute (\ref{(30)}) into (\ref{(28)}) and get
   \begin{gather}\label{(31)}
   g\frac{w_{0,0}}{w_{1,0}}=c(w_{1,0},g)+w_{0,1}.
   \end{gather}

   Substitution of (\ref{(30)}) into (\ref{(31)}) implies that
   $c(w_{0,0},w_{0,1})w = c(w_{1,0},g)w_{1,0}$,
   or the same,
   $c(w_{0,0},w_{0,1})w = D_n(c(w_{0,0},w_{0,1})w_{0,0})$.
Suppose that equation $w_{1,1}=g(w,w_{1,0},w_{0,1})$ does not admit an $m$-integral of the f\/irst order, then $c(w_{0,0},w_{0,1})w_{0,0}=D_n(c(w_{0,0},w_{0,1})w_{0,0})=C=\operatorname{const}$. Thus,
 $c(w_{0,0},w_{0,1})=C/w_{0,0}$. Finally, $g(w_{0,0},w_{1,0},w_{0,1})=\frac{w_{0,1}w_{1,0}}{w_{0,0}}+\frac{C}{w_{0,0}}$. Therefore, the equation~(\ref{discrete}) searched with $m$-integral  $F=e^{v_{1,0}-v_{0,0}}+e^{v_{1,0}-v_{2,0}}$ becomes
$e^{-v_{1,1}-v_{0,0}}=C+e^{-v_{1,0}-v_{0,1}}$, where $C$ is an arbitrary constant.
Note that this equation is symmetric with respect to variables $v_{1,0}$ and $v_{0,1}$.
Therefore, $n$-integral for the equation can be obtained by simply changing in $m$-integral
variables $v_{j,0}$ into variables $v_{0,j}$, $j=1,2$.

\subsection[Application of the algorithm of discretization to the system corresponding to the algebra $B_2$]{Application of the algorithm of discretization\\ to the system corresponding to the algebra $\boldsymbol{B_2}$}

Apply the reasonings above to the system (\ref{cont-system}) corresponding to the algebra~$B_2$. The f\/irst step has already been done in~\cite{HZY}, where the dif\/ferential-dif\/ference system was found
\begin{gather}\label{dis-cont-systemB2}
u_{1,x}^1-u_{x}^1 =e^{u^1+u^1_1-u^2_1},\qquad
u_{1,x}^2-u_{x}^2 =e^{-2u^1+u^2+u^2_1},
\end{gather}
admitting two independent $n$-integrals having the form
\begin{gather*}
I_{(1)}=2u^1_{xx}+u^2_{xx}-2\big(u^1_{x}\big)^2+2u^1_{x}u^2_{x}-\big(u^2_{xx}\big)^2,\\
I_{(2)}=u^1_{xxxx}+u^1_{x}\big(u^2_{xxx}-2u^1_{xxx}\big)+u^1_{xx}\big(4u^1_{x}u^2_{x}-2\big(u^1_{x}\big)^2-\big(u^2_{x}\big)^2\big) \\
\hphantom{I_{(2)}=}{}
 +u^1_{xx}\big(u^2_{xx}-u^1_{xx}\big)+u^2_{xx}u^1_{x}\big(u^1_{x}-2u^2_{x}\big)+\big(u^1_{x}\big)^4+\big(u^1_{x}\big)^2\big(u^2_{x}\big)^2
 -2\big(u^1_{x}\big)^2u^2_{x},
\end{gather*}
which are also integrals of fully continuous system (\ref{cont-system}) corresponding to the same algebra. It was shown in \cite{HZY} that system (\ref{dis-cont-systemB2}) admits also $x$-integrals
\begin{gather}
F_{(1)}=e^{-u^1_0+u^1_1} + e^{-u^1_1+u^1_2+u^2_2-u^2_{3}} + e^{u^1_1-u^1_2-u^2_1+u^2_2} + e^{u^1_2-u^1_{3}},\nonumber\\
F_{(2)}= e^{-u^2_0+u^2_1} + e^{-2u^1_0+2u^1_1+u^2_1-u^2_2} + 2e^{-u^1_0+2u^1_1-u^1_2} + e^{2u^1_1-2u^1_2-u^2_1+u^2_2} + e^{u^2_2-u^2_{3}}.\label{integral_B_2}
\end{gather}

The goal of this section is to construct a dif\/ference-dif\/ference system having the same func\-tions~$F_{(1)}$ and $F_{(2)}$ as their $m$-integrals.

To make the formulas shorter we change the variables
\begin{gather}\label{zam}
a=e^{-u^1},\qquad b=e^{-u^2}
\end{gather} where $a$, $b$ are new unknowns.

The given $m$-integrals in these variables read
\begin{gather}\label{fi_c2}
F_{(1)}=\frac{a_0}{a_1} +\frac{a_1b_3}{a_2b_2}+\frac{a_2b_1}{a_1b_2}+\frac{a_3}{a_2},\qquad
F_{(2)}= \frac {b_0}{b_1}+\frac {{a_0}^2b_2}{{a_1}^2
b_1}+2\frac {a_0a_2}{{a_1}^2}+\frac {{a_2}
^2b_1}{a_1^2b_2}+\frac {b_3}{b_2}.
\end{gather}
Formulas (\ref{integral_B_2}) def\/ine integrals for a semi-discrete system~(\ref{dis-cont-systemB2}) while~(\ref{fi_c2}) def\/ines integrals for a certain dif\/ference-dif\/ference system, that is why the variables in~(\ref{fi_c2}) should be labeled by a double index, however we omitted here and below in this section the second index, because its value is zero for all considered variables. Now we will look for the equations desired in the variables $a$, $b$ in the form
\begin{gather*}
a_{1,1}=f(a_{0,0},a_{1,0},a_{0,1},b_{0,0},b_{1,0},b_{0,1}),\qquad
b_{1,1}=g(a_{0,0},a_{1,0},a_{0,1},b_{0,0},b_{1,0},b_{0,1}).
\end{gather*}
Substitute the integrals (\ref{fi_c2}) into the equations for $m$-integrals
$D_m F_{(1)}=F_{(1)}$, $D_m F_{(2)}=F_{(2)}$ and bring them to the following form
\begin{subequations}\label{subs_fi_eqd}
\begin{gather}
\label{subs_fi_eqd_1}
\frac{a_{0,1}}{f} +\frac{fD^2_ng}{D_nf D_ng}+\frac{gD_nf}{fD_ng}+\frac{D^2_nf}{D_nf}=\frac{a_0}{a_1} +\frac{a_1b_3}{a_2b_2}+\frac{a_2b_1}{a_1b_2}+\frac{a_3}{a_2},\\
\label{subs_fi_eqd_2} \frac {b_{0,1}}{g}\!+\frac {{a_{0,1}}^2D_ng}{{f}^2
g}+2\frac {a_{0,1}D_nf}{{f}^2}\! +\frac {gD_nf
^2}{f^2D_ng}\!+\frac {D^2_ng}{D_ng}=\frac {b_0}{b_1}\!+\frac {{a_0}^2b_2}{{a_1}^2
b_1}+2\frac {a_0a_2}{{a_1}^2}\!+\frac {{a_2}
^2b_1}{a_1^2b_2}+\frac {b_3}{b_2}.\!\!\!
\end{gather}
\end{subequations}
 By dif\/ferentiating equation \eqref{subs_fi_eqd_2} with respect to the variables $a_{3}$, $b_{3}$ we get  equations
\begin{gather*}
\frac{1}{D_ng}D_n^2\frac{\partial g}{\partial a_1}=0 \ \Leftrightarrow \ \frac{\partial g}{\partial a_1}=0 ,\qquad
\frac{1}{D_ng}D_n^2\frac{\partial g}{\partial b_1}=\frac{1}{b_2} \ \Leftrightarrow \ \frac{\partial g}{\partial b_1}=\frac{b_{0,1}}{b_0},
 \end{gather*}
 which imply immediately
\[
g(a_{0,0},a_{1,0},a_{0,1},b_{0,0},b_{1,0},b_{0,1})=\frac{b_{1,0}b_{0,1}}{b_{0,0}}+\frac{g_1(a_{0,0},a_{0,1},b_{0,0},b_{0,1})}{b_{0,0}}.
\]
Similarly, dif\/ferentiation of \eqref{subs_fi_eqd_1} with respect to $a_{3}$, $b_{3}$ yields
\begin{gather*}
\frac{1}{D_nf}D^2_n\frac{\partial f}{\partial a_1}=\frac{1}{a_2} \ \Leftrightarrow \ \frac{\partial f}{\partial a_1}=\frac{a_{0,1}}{a_0}, \\
\frac{fD_ng}{b_2D_nfD_ng}+\frac{1}{D_nf}D_n^2\frac{\partial f}{\partial b_1}=\frac{a_1}{a_2b_2} \ \Leftrightarrow \ \frac{\partial f}{\partial b_1}=\frac{1}{a_0b_0}D_n^{-1}(a_0f-a_1a_{0,1}).
\end{gather*}
From these equations we get
\[
f(a_{0,0},a_{1,0},a_{0,1},b_{0,0},b_{1,0},b_{0,1})=\frac{a_{1,0}a_{0,1}}{a_{0,0}}+\frac{C b_{1,0}}{a_{0,0}}.
\]
By dif\/ferentiating~\eqref{subs_fi_eqd} with respect to $a_{2}$ we obtain
\begin{gather*}
\frac{a_{1,1}}{b_2b_{2,1}}\frac{\partial}{\partial a_2}\frac{D^2_ng_1}{a_{2,1}}+\frac{b_{1,1}}{a_{1}b_{2,1}}=\frac{b_1}{a_1b_2},\\
2\frac{a_{0,1}}{a_{1,1}a_{1}}+2\frac{a_{2,1}b_{1,1}}{a_1a_{1,1}b_{2,1}}+\frac1{b_{2,1}b_{2,0}}\frac{\partial}{\partial a_2}D^2_ng_1=2\frac{a_0}{a_1}+2\frac{a_2b_1}{a_1^2b_2}.
\end{gather*}
The last two equations can be rewritten as follows
\begin{gather*}
\frac{\partial}{\partial a_1}\frac{D_ng_1}{a_{1,1}}=\frac{g_1}{a_0a_{0,1}},\qquad
\frac{\partial}{\partial a_1}D_ng_1=2\frac{a_{1,1}g_1}{a_{0,1}a_0}.
\end{gather*}
By getting rid of the variable $\partial_{a_1}D_ng_1$ in these equations we f\/ind
\[
\frac{D_ng_1}{a_{1,1}^2}=\frac{g_1}{a_{0,1}^2} \ \Leftrightarrow \ g_1=C_1a_{0,1}^2.
\]
The last implication is due to the assumption that $F_{(1)}$ and $F_{(2)}$ are integrals of the lowest order.

As a result we get equations
\begin{gather*}
a_{1,1}a_{0,0}-a_{1,0}a_{0,1}=Cb_{1,0},\qquad b_{1,1}b_{0,0}-b_{1,0}b_{0,1}=C_1a_{0,1}^2.
\end{gather*}
After some rescaling we can put $C=C_1=1.$ Finally, we end up with the system of discrete equations
\begin{gather*}
a_{1,1}a_{0,0}-a_{1,0}a_{0,1}=b_{1,0},\qquad b_{1,1}b_{0,0}-b_{1,0}b_{0,1}=a_{0,1}^2.
\end{gather*}
corresponding to the algebra $B_2$. Turn back to the original variables $u^1$, $u^2$ (see~(\ref{zam}))
\begin{gather*}
\Delta{u^1}=e^{u^1_{0,1}+u^1_{1,0}-u^2_{1,0}},\qquad
\Delta{u^2}=e^{-2u^1_{0,1}+u^2_{0,1}+u^2_{1,0}}.
\end{gather*}
By construction the last system admits pair of integrals (\ref{integral_B_2}) in which the second index for the variables $u^1$, $u^2$ is omitted, since its value is the same for all variables. To f\/ind the integrals in the other direction one uses the discrete symmetry $n\leftrightarrow -m$ of the system.

\subsection[Integrals of the systems corresponding to the algebras $A_2$, $G_2$, $D_3$]{Integrals of the systems corresponding to the algebras $\boldsymbol{A_2}$, $\boldsymbol{G_2}$, $\boldsymbol{D_3}$}

\subsubsection[System corresponding to the algebra $A_2$]{System corresponding to the algebra $\boldsymbol{A_2}$}

The Cartan matrix of $A_2$ is
\[
A= \begin{pmatrix}2&-1\\-1&2\end{pmatrix}.
\]
The discrete system for $A_2$ looks as follows
\begin{gather}\label{delta_A_2}
\Delta\big(u^1\big)=e^{u_{0,1}^1+u_{1,0}^1-u_{1,0}^2},\qquad
\Delta\big(u^2\big)=e^{-u_{0,1}^1+u_{0,1}^2+u_{1,0}^2}.
\end{gather}
For system (\ref{delta_A_2})  $m$-integrals are
\begin{gather*}
F_{(1)}=e^{-u^2_{0,0}+u^2_{1,0}} + e^{-u^1_{0,0}+u^1_{1,0}+u^2_{1,0}-u^2_{2,0}} + e^{u^1_{1,0}-u^1_{2,0}},\\
F_{(2)}=e^{-u^1_{0,0}+u^1_{1,0}} + e^{u^1_{1,0}-u^1_{2,0}-u^2_{1,0}+u^2_{2,0}} + e^{u^2_{2,0}-u^2_{3,0}}.
\end{gather*}
To f\/ind the integrals in the other direction one uses the discrete symmetry $n\leftrightarrow -m$ of the system.

\subsubsection[System corresponding to the algebra $G_2$]{System corresponding to the algebra $\boldsymbol{G_2}$}

The Cartan matrix of $G_2$ is
\[
A= \begin{pmatrix}
2&-1\\-3&2\end{pmatrix}.
\]
The discrete system for $G_2$ is of the form
\begin{gather*}
\Delta\big(u^1\big)=e^{u_{0,1}^1+u_{1,0}^1-u_{1,0}^2},\qquad
\Delta\big(u^2\big)=e^{-3u_{0,1}^1+u_{0,1}^2+u_{1,0}^2}.
\end{gather*}
Its $m$-integrals are
\begin{gather*}
F_{(1)}=e^{u^1_{1,0}-u^1_{2,0}}+e^{u^1_{-1,0}-u^1_{-2,0}}+
e^{u^2_{1,0}+u^1_{1,0}-u^1_{1,0}-u^2_{2,0}}+e^{u^1_{-1,0}+u^2_{0,0}-u^1_{0,0}-u^2_{-1,0}} \\
\hphantom{F_{(1)}=}{}
+e^{u^2_{1,0}+2u^1_{0,0}-u^2_{0,0}-2u^1_{1,0}}+
e^{u^2_{0,0}+2u^1_{0,0}-u^2_{1,0}-2u^1_{-1,0}}+
2e^{2u^1_{0,0}-u^1_{-1,0}-u^1_{1,0}},\\
F_{(2)}=e^{u^2_{2,0}-u^2_{3,0}}+e^{3u^1_{0,0}+u^2_{0,0}-u^2_{1,0}-3u^1_{-1,0}}+3e^{u^1_{0,0}+u^2_{1,0}-u^1_{2,0}-u^2_{0,0}}
+3e^{u^1_{0,0}+u^1_{1,0}-u^1_{-1,0}-u^1_{2,0}} \\
\hphantom{F_{(2)}=}{}
+3e^{u^2_{1,0}+3u^1_{1,0}-2u^1_{0,0}-u^2_{2,0}-u^1_{2,0}}
+3e^{3u^1_{1,0}+u^1_{0,0}-2u^1_{2,0}}
+3e^{3u^1_{0,0}+u^1_{1,0}-u^1_{-1,0}} \\
\hphantom{F_{(2)}=}{}
+3e^{3u^1_{0,0}+u^2_{1,0}-u^2_{0,0}-2u^1_{1,0}-u^1_{-1,0}}
+3e^{u^2_{1,0}+u^2_{1,0}-u^2_{2,0}-u^1_{-1,0}}
+2e^{2u^2_{1,0}+u^2_{0,0}-u^2_{2,0}} \\
\hphantom{F_{(2)}=}{}
 +e^{3u^1_{1,0}+u^2_{2,0}-u^2_{1,0}-3u^1_{2,0}}\!
+e^{3u^1_{1,0}+2u^2_{1,0}-3u^1_{0,0}-2u^2_{2,0}}\!
+e^{3u^1_{0,0}+2u^2_{1,0}-2u^2_{0,0}-3u^1_{1,0}}\!
+e^{u^2_{2,0}+u^2_{-1,0}}.\!
\end{gather*}
To f\/ind the integrals in the other direction one uses the discrete symmetry $n\leftrightarrow -m$ of the system.

\subsubsection[System corresponding to the algebra $D_3$]{System corresponding to the algebra $\boldsymbol{D_3}$}

The Cartan matrix of $D_3$ is
\[
A= \begin{pmatrix}
2&-1&-1\\-1&2&0\\-1&0&2\end{pmatrix}.
\]
The discrete system for $D_3$ looks as
\begin{gather*}
\Delta\big(u^1\big)=e^{u_{0,1}^1+u_{1,0}^1-u_{0,1}^2-u_{0,1}^3},\qquad
\Delta\big(u^2\big)=e^{u_{0,1}^2+u_{1,0}^2-u_{1,0}^1},\qquad
\Delta\big(u^3\big)=e^{u_{0,1}^3+u_{1,0}^3-u_{1,0}^1}.
\end{gather*}
Its $m$-integrals are
\begin{gather*}
F_{(1)}=e^{u^3_{0,0}-u^3_{-1,0}}+e^{u^2_{1,0}-u^2_{2,0}}+e^{u^2_{1,0}+u^1_{1,0}-u^1_{2,0}-u^2_{0,0}}+e^{u^1_{1,0}+u^3_{0,0}-u^1_{0,0}-u^3_{1,0}},\\
F_{(2)}=e^{u^2_{0,0}-u^2_{-1,0}}+e^{u^3_{1,0}-u^3_{2,0}}+e^{u^3_{1,0}+u^1_{1,0}-u^1_{2,0}-u^3_{0,0}}+e^{u^1_{1,0}+u^2_{0,0}-u^1_{0,0}-u^2_{1,0}},
\\
F_{(3)}=e^{u^1_{1,0}-u^1_{2,0}}+e^{u^1_{0,0}-u^1_{-1,0}}+e^{u^2_{0,0}-u^2_{1,0}+u^3_{0,0}-u^3_{-1,0}}+e^{u^2_{0,0}-u^2_{-1,0}+u^3_{0,0}-u^3_{1,0}}+\\
\hphantom{F_{(3)}=}{}
+e^{u^1_{1,0}-u^1_{0,0}+u^2_{0,0}-u^2_{1,0}+u^3_{0,0}-u^3_{1,0}}
+e^{u^1_{0,0}-u^1_{1,0}+u^2_{0,0}-u^2_{-1,0}+u^3_{0,0}-u^3_{-1,0}}.
\end{gather*}
By applying Theorem~\ref{independent_int} one can prove that these integrals provide a complete set of independent integrals. To f\/ind the integrals in the other direction one uses the discrete symmetry $n\leftrightarrow -m$ of the system.

\section{Symmetries of discrete systems} \label{section3}

In this section we demonstrate that the discrete systems admit generalized symmetries. For the case of the simple Lie algebras the systems admit local symmetries while for the case of af\/f\/ine algebras the symmetries are nonlocal.

\subsection[Generalized symmetries for the system $A_2$]{Generalized symmetries for the system $\boldsymbol{A_2}$}

Higher symmetries for system corresponding to $A_2$ are found by using the method suggested in~\cite{GGH}  (see also~\cite{YL2}). The lowest order symmetry is of the form
\[
\frac{du^1_{0,0}}{dt}=\frac{3e^{u^1_{0,0}-u^1_{1,0}}}{D_m^{-1}F_{(1)}}-1,\qquad
\frac{du^2_{0,0}}{dt}=1-\frac{3e^{u^2_{0,0}-u^2_{-1,0}}}{D_m^{-1}F_{(1)}}.
\]
The next order symmetry depends on arbitrary functions $f(x,y)$, $g(x)$ of two and respectively one variable and has the form
\begin{gather*}
\frac{du^1_{0,0}}{dt} =\frac{f(F_{(1)},D_m^{-1}F_{(2)})+g(D_m^{-1}F_{(1)})}{D_m^{-1}F_{(2)}}e^{u^1_{0,0}-u^1_{-1,0}}
+f\big(F_{(1)},D_m^{-1}F_{(2)}\big)-g(F_{(1)}) \\
\hphantom{\frac{du^1_{0,0}}{dt} =}{}
+g\big(D_m^{-1}F_{(1)}\big)-\frac{f(F_{(1)},D_m^{-1}F_{(2)})+f(D_m^{-2}F_{(1)},D_m^{-2}F_{(2)})-g(F_{(1)})}{D_m^{-1}F_{(1)}}e^{u^1_{0,0}-u^1_{1,0}},\\
\frac{du^2_{0,0}}{dt} =\frac{f(F_{(1)},D_m^{-1}F_{(2)})+f(D_m^{-2}F_{(1)},D_m^{-2}F_{(2)})-g(F_{(1)})}{D_m^{-1}F_{(1)}} e^{u^2_{0,0}-u^2_{-1,0}}\\
\hphantom{\frac{du^2_{0,0}}{dt} =}{}
-f\big(D_m^{-2}F_{(1)},D_m^{-2}F_{(2)}\big)+\frac{f(D_m^{-1}F_{(1)},D_m^{-2}F_{(2)})+g(D_m^{-2}F_{(1)})}{D_m^{-2}F_{(2)}}e^{u^2_{0,0}-u^2_{1,0}}.
\end{gather*}

\subsection[Evaluation of hyperbolic type symmetries for the system $A^{(1)}_1$]{Evaluation of hyperbolic type symmetries for the system $\boldsymbol{A^{(1)}_1}$}

 The method for searching nonlocal symmetries is illustrated with the following example
\begin{gather}\label{disa22}
e^{-u_{1,1}-u_{0,0}}-e^{-u_{1,0}-u_{0,1}}=e^{-2v_{1,0}},\qquad e^{-v_{1,1}-v_{0,0}}-e^{-v_{1,0}-v_{0,1}}=e^{-2u_{0,1}},
 \end{gather}
 corresponding to the algebra $A^{(1)}_1$ with the Cartan matrix
\[
A= \begin{pmatrix}2&-2\\-2&2\end{pmatrix}.
\]
Recall that the continuous version of the system is
\begin{gather}\label{difa22}
u_{xy}=\exp(2u-2v),\qquad v_{xy}=\exp(2v-2u).
\end{gather}
It can be proved that system \eqref{difa22} does not have any local generalized symmetry, however it has nonlocal generalized symmetries, the simplest one can be represented in terms of $s=u+v$, $r=u-v$ as follows
\begin{gather*}
r_t=r_{xxx}-2r_x^3,\qquad s_{xt}=2r_xr_{xxx}-3r_x^4-r_{xx}^2+F(W,W_x).
\end{gather*}
Here $F$ is an arbitrary function and $W=s_{xx}-r_x^2$ is the $y$-integral of system~\eqref{difa22}. Note that years ago in~\cite{DrS} it was observed that equation $r_t=r_{xxx}-2r_x^3$ is consistent with the system (\ref{difa22}).

For the fully discrete analogue \eqref{disa22} of the system~\eqref{difa22} we have a very similar situation. The system~\eqref{disa22} does not have any local generalized symmetry. One has to look for a nonlocal symmetry. The main trouble arising in this case is connected with the guessing of the structure of non-locality and the form of the symmetry.

Rewrite system \eqref{disa22} in the form
\begin{gather}
\label{disa22ab}a_{1,1}a_{0,0}-a_{1,0}a_{0,1}=b_{1,0}^2,\qquad b_{1,1}b_{0,0}-b_{1,0}b_{0,1}=a_{0,1}^2,
\end{gather}
where $a_{i,j}=\exp({-u_{i,j}})$, $b_{i,j}=\exp({-v_{i,j}})$.

We will look for the hyperbolic type symmetry (see Def\/inition \ref{sh}) of \eqref{disa22ab} in such a form
\begin{gather}
\partial_t a_{1,0}=f_1(a_{0,0},a_{1,0})\partial_t a_{0,0}+f(a_{-1,0},a_{0,0},a_{1,0},a_{2,0},b_{-1,0},b_{0,0},b_{1,0},b_{2,0}),\nonumber\\
\partial_t b_{1,0}=g_1(b_{0,0},b_{1,0})\partial_t b_{0,0}+g(a_{-1,0},a_{0,0},a_{1,0},a_{2,0},b_{-1,0},b_{0,0},b_{1,0},b_{2,0}).\label{fsa22}
\end{gather}
Here $f_1$, $g_1$, $f$, $g$ are unknown functions. Right hand side of~\eqref{fsa22} corresponds to the function~$G$ in~\eqref{hs}. The function $\widetilde G$ is evaluated in terms of~$G$ below by means of the compatibility conditions.

From \eqref{fsa22} we can f\/ind
\begin{gather}\label{fsa221}
\partial_t a_{1,1}=D_m (f_1)\partial_t a_{0,1}+D_m f,\qquad \partial_t b_{1,1}=D_m (g_1)\partial_t b_{0,1}+D_m g.
\end{gather}
After dif\/ferentiation \eqref{disa22} with respect to $t$ by means of the systems (\ref{fsa22}), (\ref{fsa221}) we get
\begin{gather}
\partial_t a_{0,0}(a_{1,1}-a_{0,1}f_1)+\partial_t a_{0,1}(a_{0,0}D_m f_1-a_{1,0})-2\partial_t b_{0,0}b_{1,0}g_1=\cdots,\nonumber\\
-2a_{0,1}\partial_t a_{0,1}+\partial_t b_{0,0}(b_{1,1}-b_{0,1}g_1)+\partial_t b_{0,1}(b_{0,0}D_m g_1-b_{1,0})=\cdots.\label{ur1}
\end{gather}
 Here the right hand sides do not depend on the derivatives of dynamical variables with respect to~$t$. By applying the operator $D_n$ to both sides of~\eqref{ur1} we obtain
\begin{gather}
\partial_t a_{0,0}f_1D_n(a_{1,1}-a_{0,1}f_1)+\partial_t a_{0,1}D_m f_1D_n(a_{0,0}D_m f_1-a_{1,0})-2\partial_t b_{1,0}b_{2,0}g_1D_ng_1=\cdots,\nonumber\\ -2a_{1,1}\partial_t D_m f_1 +\partial_t b_{0,0}g_1 D_n(b_{1,1}-b_{0,1}g_1)+\partial_t b_{0,1}D_m g_1 D_n(b_{0,0}D_m g_1-b_{1,0})=\cdots.\label{ur2}
\end{gather}
System of equations (\ref{ur1}), (\ref{ur2}) is a  system of linear algebraic equations with unknowns $\partial_t a_{0,0}$, $\partial_t b_{0,0}$, $\partial_t a_{0,1}$, $\partial_t b_{0,1}$. If the determinant of this system is dif\/ferent from  zero, then due to the Cramer's rule the system has unique solution, and therefore the searched generalized symmetry is local. This output is in contradiction with our previous study proving the absence of local symmetries. Thus the determinant should be zero
\begin{gather*}\left|\begin{array}{@{}cc}
a_{1,1}-f_1a_{0,1}&a_{0,0}D_m f_1-a_{1,0}\\
f_1(a_{2,1}-a_{1,1}D_nf_1)& (a_{1,0}D_nD_m f_1-a_{2,0}) D_m f_1\\
0& -2a_{1,1}D_m f_1\\
0&-2a_{0,1}
 \end{array} \right.\\
\left. \hspace*{65mm}\begin{array}{cc@{}}
  -2b_{1,0}g_1&0\\
   -2b_{2,0}g_1D_ng_1&0\\
  g_1 (b_{2,1}-b_{1,1}D_ng_1)&D_m g_1 (b_{1,0}D_n D_m g_1-b_{2,0})\\
 b_{1,1}-g_1b_{0,1}& b_{0,0}D_m g_1-b_{1,0}
\end{array}
 \right|.
\end{gather*}
Vanishing of the determinant implies
\begin{gather}
f_1(a_{0,0},a_{1,0})=a_{1,0}/a_{0,0},\qquad g_1(b_{0,0},b_{1,0})=b_{1,0}/b_{0,0},\label{f1g1}
\end{gather}
In virtue of \eqref{f1g1} equations (\ref{ur1}), (\ref{ur2}) take the form
\begin{gather}
\frac{\partial_t a_{0,0}}{a_{0,0}}+\frac{\partial_t a_{0,1}}{a_{0,1}}-2\frac{\partial_t b_{0,0}}{b_{0,0}}=\frac2{b_{1,0}}g+\frac{a_{0,1}}{b_{1,0}^2}f-\frac{a_{0,0}}{b_{1,0}^2}D_m f,\nonumber\\
\frac{\partial_t a_{0,0}}{a_{0,0}}+\frac{\partial_t a_{0,1}}{a_{0,1}}-2\frac{\partial_t b_{0,0}}{b_{0,0}}=
\frac2{b_{2,0}}D_ng+\frac{a_{1,1}}{b_{2,0}^2}D_nf-\frac{a_{1,0}}{b_{2,0}^2}D_n D_m f-\frac{f}{a_{1,0}}-\frac{D_mf}{a_{1,1}}+\frac2{b_{1,0}}g,
\nonumber\\
\frac{\partial_t b_{0,0}}{b_{0,0}}+\frac{\partial_t b_{0,1}}{b_{0,1}}-2\frac{\partial_t a_{0,1}}{a_{0,1}}=\frac{b_{0,1}}{a_{0,1}^2}g+\frac{b_{0,0}}{a_{0,1}^2}D_m g,\nonumber\\
\frac{\partial_t b_{0,0}}{b_{0,0}}+\frac{\partial_t b_{0,1}}{b_{0,1}}-2\frac{\partial_t a_{0,1}}{a_{0,1}}=\frac{b_{1,1}}{a_{1,1}^2}D_ng+\frac{b_{1,0}}{a_{1,1}^2}D_n D_m g+\frac{2}{a_{1,1}} D_m f-\frac{g}{b_{1,0}}-\frac{D_m g}{b_{1,1}}.
\label{one_com}
\end{gather}
We can see that the left hand sides of the f\/irst and third equations as well as the left hand sides of the second and fourth equations of the last system are identical. Comparison of their right hand sides gives a system of two equation for~$f$,~$g$
\begin{gather}
\label{lin1}\frac{a_{1,0}}{b_{2,0}^2}D_n D_m f-\frac{a_{1,1}}{b_{2,0}^2}D_nf-\frac{a_{1,0}a_{0,1}}{a_{1,1}b_{1,0}^2} D_mf+\frac{a_{1,1}a_{0,0}}{a_{1,0}b_{1,0}^2}f-\frac2{b_{2,0}}D_ng=0,\\
\label{lin2}\frac{b_{1,0}}{a_{1,1}^2}D_n D_m g+\frac{b_{1,1}}{a_{1,1}^2}D_ng -\frac{b_{1,0}b_{0,1}}{b_{1,1}a_{0,1}^2} D_m g-\frac{b_{1,1}b_{0,0}}{b_{1,0}a_{0,1}^2}g+\frac{2}{a_{1,1}} D_m f=0.
\end{gather}

Dif\/ferentiate \eqref{lin1} and \eqref{lin2} with respect to $a_{3,0}$, $b_{3,0}$, $b_{-1,0}$ or $a_{-1,0}$ and get
\begin{gather}
\frac{b_{0,0}a_{0,0}}{a_{1,0}^2}\frac{\partial f}{\partial b_{-1,0}}= D_m\left(\frac{b_{0,0}a_{0,0}}{a_{1,0}^2} \frac{\partial f}{\partial b_{-1,0}}\right),\nonumber\\
\label{dif_kr1}
D_m \left(\frac{a_{0,0}^2}{a_{1,0}^2}\frac{\partial f}{\partial a_{-1,0}}\right)=\frac{a_{0,0}^2}{a_{1,0}^2}\frac{\partial f}{\partial a_{-1,0}}+\frac{2a_{-1,1}a_{0,1}^2}{a_{1,1}^2b_{0,0}}D_m\frac{\partial f}{\partial b_{-1,0}}, \\
\frac{a_{1,0}}{b_{0,0}}\frac{\partial g}{\partial a_{2,0}}=\frac{a_{1,1}}{b_{0,1}}D_m \frac{\partial g}{\partial a_{2,0}},\qquad
D_m\frac{\partial g}{\partial b_{2,0}}+\frac{2b_{1,0}b_{2,0}}{a_{1,0}b_{1,1}} D_m\frac{\partial g}{\partial a_{2,0}}=\frac{b_{1,0}b_{0,1}}{b_{0,0}b_{1,1}}\frac{\partial g}{\partial b_{2,0}},\label{dif_kr2}\\
\frac{\partial f}{\partial a_{2,0}}+\frac{2b_{1,0}}{a_{0,1}}\frac{\partial g}{\partial a_{2,0}}=\frac{a_{1,1}a_{0,0}}{a_{1,0}a_{0,1}} D_m \frac{\partial f}{\partial a_{2,0}},\nonumber\\
\frac{\partial f}{\partial b_{2,0}}+\frac{2b_{1,0}}{a_{0,1}}\frac{\partial g}{\partial b_{2,0}}=\frac{2a_{0,0}b_{2,0}}{a_{1,0}a_{0,1}} D_m\frac{\partial f}{\partial a_{2,0}}+\frac{a_{0,0}b_{1,1}}{a_{0,1}b_{1,0}}D_m\frac{\partial f}{\partial b_{2,0}},\nonumber\\
\frac{\partial g}{\partial b_{-1,0}}+\frac{2b_{1,0}a_{0,0}^2}{a_{1,0}b_{1,-1}b_{0,0}}\frac{\partial f}{\partial b_{-1,0}}=\frac{b_{1,0}^2b_{0,-1}^2}{b_{1,-1}^2b_{00,}^2} D_m^{-1}\frac{\partial g}{\partial b_{-1,0}},\nonumber\\
\frac{\partial g}{\partial a_{-1,0}}+\frac{2a_{0,0}^2b_{1,0}}{a_{1,0}b_{1,-1}b_{0,0}}\frac{\partial f}{\partial a_{-1,0}}=\frac{b_{1,0}^2a_{0,-1}b_{0,-1}}{a_{0,0}b_{0,0}b_{1,-1}}D_m^{-1}\frac{\partial g}{\partial a_{-1,0}}\nonumber\\
\hphantom{\frac{\partial g}{\partial a_{-1,0}}+\frac{2a_{0,0}^2b_{1,0}}{a_{1,0}b_{1,-1}b_{0,0}}\frac{\partial f}{\partial a_{-1,0}}=}{}
+\frac{4a_{0,0}^2a_{-1,0}b_{1,0}}{a_{1,0}b_{0,-1}b_{1,-1}b_{0,0}}\frac{\partial f}{\partial b_{-1,0}}+\frac{2a_{-1,0}}{b_{0,-1}}\frac{\partial g}{\partial b_{-1,0}}\nonumber.
\end{gather}
We use the last system of eight equations in order to specify dependence of the functions $f$, $g$ on the variables $a_{-1,0}$, $a_{2,0}$, $b_{-1,0}$, $b_{2,0}$ corresponding to the lowest and highest values of the f\/irst index.
From the f\/irst equation in~\eqref{dif_kr1} we get
\begin{gather*}
\frac{\partial f}{\partial b_{-1,0}}=\frac{C_7a_{1,0}^2}{b_{0,0}a_{0,0}}.
\end{gather*}
Then the second equation takes the form
\begin{gather*}
D_m \left(\frac{a_{0,0}^2}{a_{1,0}^2}\frac{\partial f}{\partial a_{-1,0}}\right)=\frac{a_{0,0}^2}{a_{1,0}^2}\frac{\partial f}{\partial a_{-1,0}}+\frac{2C_7a_{-1,1}a_{0,1}}{b_{0,1}b_{0,0}}.
\end{gather*}
It can easily be proved that the expression $\frac{2a_{-1,1}a_{0,1}}{b_{0,1}b_{0,0}}$ does not belong to the image of the operator $D_m -1$, so we have to put $C_7=0$ and then
\begin{gather*}
\frac{\partial f}{\partial a_{-1,0}}=\frac{C_4a_{1,0}^2}{a_{0,0}^2}.
\end{gather*}
From the pair of equations \eqref{dif_kr2} we have
\begin{gather*}
\frac{\partial g}{\partial a_{2,0}}=0,\qquad \frac{\partial g}{\partial b_{2,0}}=\frac{C_2b_{0,0}}{b_{1,0}}.
\end{gather*}
Continuing this way one can f\/ind dependence of the functions $f$, $g$ on the variables $a_{-1,0}$, $a_{2,0}$, $b_{-1,0}$, $b_{2,0}$
\begin{gather*}
f=\frac{a_{0,0}b_{2,0}}{b_{1,0}}C_1+C_2\left(\frac{a_{0,0}a_{2,0}}{a_{1,0}}+
\frac{a_{0,0}b_{2,0}^2}{a_{1,0}b_{1,0}^2}+\frac{2b_{0,0}a_{1,0}b_{2,0}}{b_{1,0}^2}\right)\\
\hphantom{f=}{} +
\frac{a_{1,0}^2a_{-1,0}}{a_{0,0}^2}C_4+f_1(a_{0,0},a_{1,0},b_{0,0},b_{1,0}),\nonumber\\
g=\frac{b_{0,0}b_{2,0}}{b_{1,0}}C_2+C_4\left(\frac{b_{1,0}^2b_{-1,0}}{b_{0,0}^2}+
\frac{b_{1,0}^3a_{-1,0}^2}{a_{0,0}^2b_{0,0}^2}
+\frac{2b_{1,0}a_{1,0}a_{-1,0}}{a_{0,0}^2}\right)\\
\hphantom{g=}{} +
\frac{b_{1,0}^2a_{-1,0}}{a_{0,0}b_{0,0}}C_5+g_1(a_{0,0},a_{1,0},b_{0,0},b_{1,0}).\nonumber
\end{gather*}
For the functions $f_1$, $g_1$ from the same system (\ref{lin1}), (\ref{lin2}) we obtain the following non-homogeneous equations
\begin{gather}
\label{lin3}\frac{a_{1,0}}{b_{2,0}^2}D_nD_m f_1-\frac{a_{1,1}}{b_{2,0}^2}D_nf_1-\frac{a_{1,0}a_{0,1}}{a_{1,1}b_{2,0}^2} D_mf_1+\frac{a_{1,1}a_{0,0}}{a_{1,0}b_{1,0}^2}f_1-\frac2{b_{2,0}}D_ng_1=\cdots,\\
\label{lin4}
\frac{b_{1,0}}{a_{1,1}^2}D_n D_m g_1+\frac{b_{1,1}}{a_{1,1}^2}D_ng_1 -\frac{b_{1,0}b_{0,1}}{b_{1,1}a_{0,1}^2}D_m g_1-\frac{b_{1,1}b_{0,0}}{b_{1,0}a_{0,1}^2}g_1+\frac{2}{a_{1,1}}D_m f_1=\cdots.
\end{gather}
Dif\/ferentiating \eqref{lin4} with respect to the variable $a_{2,0}$ we get
\begin{gather*}
\frac{a_{0,0}}{b_{0,0}}\frac{\partial g_1}{\partial a_{1,0}}-D_m \left(\frac{a_{0,0}}{b_{0,0}}\frac{\partial g_1}{\partial a_{1,0}}\right)=\frac{2C_2(a_{0,1}^3a_{1,0}-b_{1,0}^3b_{0,1})}{a_{0,0}a_{0,1}b_{1,1}b_{1,0}}
+2(C_4-C_2)\frac{a_{0,1}^3a_{-1,0}-b_{1,1}b_{0,0}^3}{a_{0,0}b_{0,0}a_{0,1}b_{0,1}}.
\end{gather*}
 From this equation we can f\/ind
\begin{gather*}
C_4=C_2,\qquad \frac{a_{0,0}}{b_{0,0}}\frac{\partial g_1}{\partial a_{1,0}}=\frac{2C_2a_{1,0}b_{0,0}}{a_{0,0}b_{1,0}}+C_6,
 \end{gather*}
 therefore
\begin{gather*}
g_1=C_2\frac{a_{1,0}^2b_{0,0}^2}{a_{0,0}^2b_{1,0}}+C_6\frac{a_{1,0}b_{0,0}}{a_{0,0}}+g_2(a_{0,0},b
_{0,0},b_{1,0}).
\end{gather*}
Dif\/ferentiating \eqref{lin4} with respect to $b_{2,0}$ we get
\begin{gather*}
D_m \left(\frac{\partial g_2}{\partial b_{1,0}}\right)-\frac{\partial g_2}{\partial b_{1,0}}=2(C_1-C_6)\frac{b_{0,0}b_{1,0}}{a_{0,0}a_{0,1}}
+2(C_1-C_5)\frac{a_{0,1}^2a_{0,0}a_{-1,1}-
b_{0,0}^2b_{1,0}b_{0,1}}{a_{0,0}b_{0,0}a_{0,1}b_{0,1}},
\end{gather*}
therefore
\begin{gather*}
C_6=C_1,\qquad C_5=C_1,\qquad g_2=C_3b_{1,0}+g_3(a_{0,0},b
_{0,0}).
\end{gather*}
Now apply the operator $D^{-1}_n$ to the equation \eqref{lin3} and then dif\/ferentiate the obtained result with respect to the variable $b_{-1,0}$
\begin{gather*}
D_m \left(\frac{a_{0,0}b_{1,0}}{a_{1,0}^2}\frac{\partial f_1}{\partial b_{0,0}}\right)-\frac{a_{0,0}b_{1,0}}{a_{1,0}^2}\frac{\partial f_1}{\partial b_{0,0}}=2C_2\frac{b_{1,0}^3b_{0,1}-a_{1,0}a_{0,1}^3}{a_{0,0}a_{0,1}b_{1,0}b_{1,1}},
\end{gather*}
therefore
\begin{gather*}
f_1=C_2\frac{a_{1,0}^3b_{0,0}^2}{a_{0,0}^2b_{1,0}^2}+C_8\frac{a_{1,0}^2b_{0,0}}{a_{0,0}b_{1,0}}+f_2(a_{0,0},a_{1,0},b_{1,0}).
\end{gather*}
Substitute in \eqref{lin3} the expression found instead of $f_1$ and dif\/ferentiate with respect to $a_{2,0}$ to get
\begin{gather*}
D_m \left(\frac{\partial f_2}{\partial a_{1,0}}\right)-\frac{\partial f_2}{\partial a_{1,0}}=2(C_1-C_8)\frac{a_{1,0}a_{0,1}^3-
b_{1,0}^3b_{0,1}}{a_{0,0}a_{0,1}b_{1,0}b_{1,1}},
\end{gather*}
 therefore
\begin{gather*}
C_8=C_1,\qquad f_2=C_9 a_{1,0}+f_3(a_{0,0},b_{1,0}).
\end{gather*}
For the functions $f_3$, $g_3$ we have a system
\begin{gather}
\frac{a_{1,0}}{b_{2,0}^2}D_n D_m f_3-\frac{a_{1,1}}{b_{2,0}^2}D_nf_3-\frac{a_{1,0}a_{0,1}}{a_{1,1}b_{2,0}^2} D_mf_3+\frac{a_{1,1}a_{0,0}}{a_{1,0}b_{1,0}^2}f_3-\frac2{b_{2,0}}D_ng_3=2(C_9-C_3),\nonumber\\
\label{lin5}
\frac{b_{1,0}}{a_{1,1}^2}D_n D_m g_3+\frac{b_{1,1}}{a_{1,1}^2}D_ng_3 -\frac{b_{1,0}b_{0,1}}{b_{1,1}a_{0,1}^2} D_m g_3-\frac{b_{1,1}b_{0,0}}{b_{1,0}a_{0,1}^2}g_3+\frac{2}{a_{1,1}}D_m f_3=2(C_9-C_3).
\end{gather}
We see that \eqref{lin5} is satisf\/ied identically if one chooses
\begin{gather*}
C_9=C_3,\qquad f_3=0,\qquad g_3=0.
\end{gather*}
Thus we f\/ind the f\/inal form of the symmetry searched
\begin{gather}
\frac{\partial_t a_{1,0}}{a_{1,0}}-\frac{\partial_t a_{0,0}}{a_{0,0}}= C_1\left(\frac{a_{0,0}b_{2,0}}{a_{1,0}b_{1,0}}+\frac{a_{1,0}b_{0,0}}{a_{0,0}b_{1,0}}\right)\nonumber\\
\hphantom{\frac{\partial_t a_{1,0}}{a_{1,0}}-\frac{\partial_t a_{0,0}}{a_{0,0}}= }{}
+C_2\left(\frac{a_{1,0}a_{-1,0}}{a_{0,0}^2}+\frac{2b_{0,0}b_{2,0}}{b_{1,0}^2}
+\frac{a_{1,0}^2b_{0,0}^2}{a_{0,0}^2b_{1,0}^2}+\frac{a_{0,0}a_{2,0}}{a_{1,0}^2}+\frac{a_{0,0}^2b_{2,0}^2}{a_{1,0}^2b_{1,0}^2}\right)+C_3,
\nonumber\\
\frac{\partial_t b_{1,0}}{b_{1,0}}-\frac{\partial_t b_{0,0}}{b_{0,0}}= C_1\left(\frac{a_{-1,0}b_{1,0}}{a_{0,0}b_{0,0}}+\frac{a_{1,0}b_{0,0}}{a_{0,0}b_{1,0}}\right)\label{fun_G}\\
\hphantom{\frac{\partial_t b_{1,0}}{b_{1,0}}-\frac{\partial_t b_{0,0}}{b_{0,0}}= }{}
+ C_2\left(\frac{2a_{1,0}a_{-1,0}}{a_{0,0}^2}+\frac{b_{0,0}b_{2,0}}{b_{1,0}^2}
+\frac{a_{1,0}^2b_{0,0}^2}{a_{0,0}^2b_{1,0}^2}+\frac{b_{-1,0}b_{1,0}}{b_{0,0}^2}+\frac{a_{-1,0}b_{1,0}^2}{a_{0,0}^2b_{0,0}^2}\right)+C_3.
\nonumber
\end{gather}
Here $C_1$, $C_2$, $C_3$ are arbitrary constants. From \eqref{one_com} one can f\/ind the function $\widetilde G$, then \eqref{hs} looks like
\begin{gather}
\frac{\partial_t a_{0,1}}{a_{0,1}}= \frac{\partial_t b_{0,0}}{b_{0,0}}-\frac{\partial_t a_{0,0}}{a_{0,0}}+C_1\frac{b_{1,0}}{a_{0,0}}\left(\frac{2a_{-1,0}}{b_{0,0}}-\frac{b_{0,0}}{a_{0,1}}\right)\nonumber\\
\hphantom{\frac{\partial_t a_{0,1}}{a_{0,1}}=}{}
+ C_2\left(\frac{2b_{1,0}b_{-1,0}}{b_{0,0}^2}+\frac{2b_{1,0}^2a_{-1,0}^2}{a_{0,0}^2b_{0,0}^2}
+\frac{2a_{1,0}a_{-1,0}}{a_{0,0}^2}-\frac{a_{-1,0}b_{1,0}^2}{a_{0,0}^2a_{0,1}}-\frac{b_{0,0}^2a_{1,0}}{a_{0,0}^2a_{0,1}}\right)+C_3,\nonumber\\
\frac{\partial_t b_{0,1}}{b_{0,1}}= \frac{2\partial_t a_{0,0}}{a_{0,0}}-\frac{3\partial_t b_{0,0}}{b_{0,0}}+C_2\left(\frac{2b_{1,0}a_{-1,0}}{a_{0,0}b_{0,0}}+\frac{b_{0,0}a_{0,1}}{a_{0,0}b_{0,1}}
-\frac{a_{-1,0}a_{0,1}^2}{a_{0,0}b_{0,0}b_{0,1}}\right)\nonumber\\
\hphantom{\frac{\partial_t b_{0,1}}{b_{0,1}}=}{}
+ C_2\left(\frac{2a_{1,0}a_{-1,0}}{a_{0,0}^2}+\frac{2b_{1,0}b_{-1,0}}{b_{0,0}^2}
+\frac{2a_{0,1}b_{1,0}a_{-1,0}}{a_{0,0}^2b_{0,1}}+\frac{2b_{1,0}^2a_{-1,0}^2}{a_{0,0}^2b_{0,0}^2}
-\frac{b_{0,0}^2b_{1,0}}{a_{0,0}^2b_{0,1}}\right.\nonumber\\ \left.
\hphantom{\frac{\partial_t b_{0,1}}{b_{0,1}}=}{}
-\frac{a_{0,1}^2b_{-1,0}}{b_{0,0}^2b_{0,1}}-\frac{a_{0,1}^2a_{-1,0}^2b_{1,0}}{a_{0,0}^2b_{0,0}^2b_{0,1}}\right)+C_3.\label{fun_wtG}
\end{gather}
Thus we have proved the following
\begin{Theorem} Equations \eqref{fun_G} and \eqref{fun_wtG} define a hyperbolic type symmetry for equa\-tion~\eqref{disa22ab}.
\end{Theorem}

\begin{Remark}By applying the replacement $n\leftrightarrow -m$ to the equations \eqref{fun_G} and \eqref{fun_wtG} one can obtain the second hyperbolic type symmetry for equation~\eqref{disa22ab}.
\end{Remark}

By analogy with the continuous case we have a dif\/ferential constraint completely consistent with system (\ref{disa22ab}), which is obtained from the last system by applying the operator $(D_n-1)^{-1}$ to the dif\/ference of the two equations
\begin{gather*}
\frac{\partial_t b_{0,0}}{b_{0,0}}-\frac{\partial_t a_{0,0}}{a_{0,0}}=C_1\frac{a_{-1,0}b_{1,0}}{a_{0,0}b_{0,0}}+C_2\left(\frac{a_{-1,0}a_{1,0}}{a_{0,0}^2}+
\frac{b_{-1,0}b_{1,0}}{b_{0,0}^2}+\frac{a_{-1,0}^2b_{1,0}^2}{a_{0,0}^2b_{0,0}^2}\right).
\end{gather*}
Under the Cole--Hopf type transformation
\begin{gather*}
\hat a_{0,0}=\frac{a_{1,0}}{a_{0,0}},\qquad \hat b_{0,0}=\frac{b_{1,0}}{b_{0,0}}
\end{gather*}
system (\ref{disa22ab}) converts to the following one
\begin{gather}
\hat a_{1,1}\hat a_{0,0}\hat a_{0,1}-\hat a_{1,0}\hat a_{0,1}\hat a_{0,0}=\hat b_{1,0}^2(\hat a_{0,0}-\hat a_{0,1}),
 \nonumber\\
 \hat b_{1,1}\hat b_{0,0}\hat b_{0,1}-\hat b_{1,0}\hat b_{0,1}\hat b_{0,0}=\hat a_{0,1}^2(\hat b_{0,1}-\hat b_{0,0}).\label{disa22ab1}
  \end{gather}
  This transformation brings our hyperbolic type symmetry to generalized symmetry of usual form for the new system (\ref{disa22ab1})
\begin{gather*}
\partial_t \hat a_{0,0}=C_1\left(\hat b_{1,0}+\frac{\hat a_{0,0}^2}{\hat b_{0,0}}\right)+C_2\left(\frac{\hat a_{0,0}^2}{\hat a_{-1,0}}+\frac{2\hat b_{1,0}\hat a_{0,0}}{\hat b_{0,0}}+\frac{\hat a_{0,0}^3}{\hat b_{0,0}^2}+\hat a_{1,0}+\frac{\hat b_{1,0}^2}{\hat a_{0,0}}\right)+C_3\hat a_{0,0},\\
\partial_t \hat b_{0,0}=C_1\left(\frac{\hat b_{0,0}^2}{\hat a_{-1,0}}+\hat a_{0,0}\right)+C_2\left(\frac{2\hat a_{0,0}\hat b_{0,0}}{\hat a_{-1,0}}+\hat b_{1,0}+\frac{\hat a_{0,0}^2}{\hat b_{0,0}}+\frac{\hat b_{0,0}^2}{\hat b_{-1,0}}+\frac{\hat b_{0,0}^3}{\hat a_{-1,0}^2}\right)+C_3\hat b_{0,0}.
\end{gather*}

\subsection[Hyperbolic type symmetries for the system $A^{(2)}_2$]{Hyperbolic type symmetries for the system $\boldsymbol{A^{(2)}_2}$}

The Cartan matrix of $A^{(2)}_2$ is
\[
A= \begin{pmatrix}2&-1\\-4&2\end{pmatrix}.
\]
The discrete system for $A_2^{(2)}$ looks as
\begin{gather*}
\Delta\big(u^1\big)=e^{u_{0,1}^1+u_{1,0}^1-u_{1,0}^2}, \qquad
\Delta\big(u^2\big)=e^{-4u_{0,1}^1+u_{0,1}^2+u_{1,0}^2}.
\end{gather*}
First part of its hyperbolic type  symmetry (function~$G$) is
\begin{gather*}
\big(u^1_{1,0}-u^1_{0,0}\big)_t=C_1 \left(e^{2u^1_{1,0}-u^1_{0,0}-u^1_{2,0}}+
e^{2u^1_{1,0}-2u^1_{0,0}+u^2_{1,0}-u^2_{2,0}}+
e^{2u^1_{0,0}-2u^1_{1,0}+u^2_{1,0}-u^2_{0,0}}\right.\\
\left.
\hphantom{\big(u^1_{1,0}-u^1_{0,0}\big)_t=}{}
+e^{2u^1_{0,0}-u^1_{1,0}-u^1_{-1,0}}\right) +
C_2 \left(3e^{4u^1_{1,0}-2u^1_{0,0}-2u^1_{2,0}}+e^{u^1_{1,0}-u^1_{0,0}+u^1_{2,0}-u^1_{3,0}}\right.\\
\hphantom{\big(u^1_{1,0}-u^1_{0,0}\big)_t=}{}
+3e^{-u^1_{2,0}+u^2_{1,0}+u^1_{0,0}-u^2_{0,0}}\!+ 3e^{4u^1_{1,0}-u^1_{2,0}-3u^1_{0,0}+u^2_{1,0}-u^2_{2,0}}\!+2e^{u^1_{1,0}-u^1_{2,0}+u^1_{0,0}-u^1_{-1,0}}\\
\hphantom{\big(u^1_{1,0}-u^1_{0,0}\big)_t=}{}
+e^{4u^1_{1,0}-4u^1_{0,0}+2u^2_{1,0}-2u^2_{2,0}}  + 2e^{u^2_{2,0}-2u^2_{1,0}-u^2_{0,0}}+3e^{u^1_{1,0}-u^2_{2,0}+u^2_{1,0}-u^1_{-1,0}}\\
\hphantom{\big(u^1_{1,0}-u^1_{0,0}\big)_t=}{}
+e^{-u^1_{1,0}+u^1_{-1,0}+u^1_{0,0}-u^1_{-2,0}}  + e^{-4u^1_{1,0}+2u^2_{1,0}-2u^2_{0,0}+4u^1_{0,0}}\\
\hphantom{\big(u^1_{1,0}-u^1_{0,0}\big)_t=}{}
+e^{4u^1_{1,0}-u^1_{0,0}+u^2_{2,0}-3u^1_{2,0}-u^2_{1,0}}+e^{-u^1_{1,0}+u^1_{-1,0}+u^2_{0,0}-u^2_{-1,0}}
\\
\hphantom{\big(u^1_{1,0}-u^1_{0,0}\big)_t=}{}
+ 3e^{-2u^1_{1,0}-2u^1_{-1,0}+4u^1_{0,0}}+e^{-u^1_{0,0}+u^2_{2,0}+u^1_{2,0}-u^2_{3,0}}
\\
\left.\hphantom{\big(u^1_{1,0}-u^1_{0,0}\big)_t=}{}
+3e^{3u^1_{1,0}+u^2_{1,0}-u^1_{-1,0}-u^2_{0,0}+4u^1_{0,0}}+e^{-u^1_{1,0}-u^2_{1,0}-3u^1_{-1,0}+u^2_{0,0}+4u^1_{0,0}}\right)+C_3,\\
\big(u^2_{1,0}-u^2_{0,0}\big)_t=2C_1 \left(
e^{-2u^1_{1,0}+u^2_{1,0}-u^2_{0,0}+2u^1_{0,0}}\!+2e^{-u^1_{1,0}-u^1_{-1,0}+2u^1_{0,0}}\!+e^{-u^2_{1,0}+2u^1_{0,0}+u^2_{0,0}-2u^1_{-1,0}}\right)\\
\hphantom{\big(u^2_{1,0}-u^2_{0,0}\big)_t=}{}
+2C_2 \left(
2e^{u^1_{1,0}-u^1_{2,0}+u^1_{0,0}-u^1_{-1,0}}+2e^{-u^1_{2,0}+u^2_{1,0}+u^1_{0,0}-u^2_{0,0}}\right.\\
\hphantom{\big(u^2_{1,0}-u^2_{0,0}\big)_t=}{}
+4e^{-3u^1_{1,0}+u^2_{1,0}-u^1_{-1,0}-u^2_{0,0}+4u^1_{0,0}}+ e^{-4u^1_{1,0}+2u^2_{1,0}-2u^2_{0,0}+4u^1_{0,0}} \\
\hphantom{\big(u^2_{1,0}-u^2_{0,0}\big)_t=}{}
+2e^{u^1_{1,0}-u^2_{2,0}+u^2_{1,0}-u^1_{-1,0}}+e^{-u^2_{2,0}+2u^2_{1,0}-u^2_{0,0}}+
2e^{-u^1_{1,0}+u^1_{-1,0}+u^1_{0,0}-u^1_{-2,0}}
\\
\hphantom{\big(u^2_{1,0}-u^2_{0,0}\big)_t=}{}
+2e^{-u^1_{1,0}+u^1_{-1,0}+u^2_{0,0}-u^2_{-1,0}}+4e^{-u^1_{1,0}-u^2_{1,0}-3u^1_{-1,0}+u^2_{0,0}+4u^1_{0,0}}\\
\hphantom{\big(u^2_{1,0}-u^2_{0,0}\big)_t=}{}
+
2e^{-u^2_{1,0}+u^1_{0,0}+u^2_{0,0}-u^1_{-2,0}}+e^{-u^2_{1,0}+2u^2_{0,0}-u^2_{-1,0}}+6e^{-2u^1_{1,0}-2u^1_{-1,0}+4u^1_{0,0}} \\
\left.\hphantom{\big(u^2_{1,0}-u^2_{0,0}\big)_t=}{}
+
e^{-2u^2_{1,0}+4u^1_{0,0}+2u^2_{0,0}-4u^1_{-1,0}}\right)+2C_3.
\end{gather*}
Here $C_1$, $C_2$, $C_3$ are arbitrary constants. Second part of hyperbolic type symmetry (function $\widetilde G$) can be evaluated automatically from the compatibility conditions.

And for combination we have a local constraint
\begin{gather*}
\big(2u^1_{0,0}- u^2_{0,0}\big)_t=2C_1\left(e^{2u^1_{0,0}-u^1_{1,0}-u^1_{-1,0}}
+e^{2u^1_{0,0}-2u^1_{-1,0}+u^2_{0,0}-u^2_{1,0}}\right)\\
\hphantom{\big(2u^1_{0,0}- u^2_{0,0}\big)_t=}{}
+2C_2 \left(e^{-u^1_{-1,0}-u^1_{2,0}+u^1_{0,0}+u^1_{1,0}}+e^{u^1_{1,0}
+u^2_{1,0}-u^1_{-1,0}-u^2_{2,0}}\right.\\
\hphantom{\big(2u^1_{0,0}- u^2_{0,0}\big)_t=}{}
+e^{-3u^1_{1,0}+u^2_{1,0}+4u^1_{0,0}-u^1_{-1,0}-u^2_{0,0}}+
3e^{-2u^1_{1,0}+4u^1_{0,0}-2u^1_{-1,0}}\\
\hphantom{\big(2u^1_{0,0}- u^2_{0,0}\big)_t=}{}
+3e^{-u^1_{1,0}-u^2_{1,0}+4u^1_{0,0}-3u^1_{-1,0}+u^2_{0,0}}
+e^{-u^1_{1,0}+u^1_{-1,0}+u^2_{0,0}-u^2_{-1,0}} \\
\hphantom{\big(2u^1_{0,0}- u^2_{0,0}\big)_t=}{}
+
 e^{-u^1_{1,0}+u^1_{0,0}+u^1_{-1,0}-u^1_{-2,0}}+e^{-2u^2_{1,0}+4u^1_{0,0}-4u^1_{-1,0}+2u^2_{0,0}} \\
\left.
\hphantom{\big(2u^1_{0,0}- u^2_{0,0}\big)_t=}{}
+e^{-u^2_{1,0}+2u^2_{0,0}-u^2_{-1,0}}+2e^{-u^2_{1,0}+u^1_{0,0}+u^2_{0,0}-u^1_{-2,0}}\right).
\end{gather*}

\section[Characteristic $m$-algebra for the case $A_2$]{Characteristic $\boldsymbol{m}$-algebra for the case $\boldsymbol{A_2}$}\label{section4}

Let us describe brief\/ly the properties of the characteristic Lie algebras of the system
\begin{gather}\label{system1}
a_{0,0}a_{1,1}=a_{1,0}a_{0,1}+b_{1,0},\qquad
b_{0,0}b_{1,1}=b_{1,0}b_{0,1}+a_{0,1}.
\end{gather}
Recall that system (\ref{system1}) corresponds to the simple Lie algebra $A_2$. First we concentrate on the notion of the characteristic $m$-algebra for the system (\ref{system1}). Lie algebra on the other destination is studied similarly. Recall that according to the def\/inition an $m$-integral $F(a_{0,0},b_{0,0},a_{1,0},b_{1,0}$, $a_{-1,0},b_{-1,0},\dots)$ should satisfy the equation $D_mF=F$. In the coordinate representation this condition reads
\begin{gather}\label{F_1}
F(a_{0,1},b_{0,1},a_{1,1},b_{1,1},a_{-1,1},b_{-1,1},\dots)=F(a_{0,0},b_{0,0},a_{1,0},b_{1,0},a_{-1,0},b_{-1,0},\dots).
\end{gather}
Evidently the right hand side in (\ref{F_1}) does not depend on the variables $a_{0,1}$ and $b_{0,1}$ hence the conditions hold $\frac{\partial}{\partial a_{0,1}}D_m^{-1}F=0$, $\frac{\partial}{\partial b_{0,1}}D_m^{-1}F=0$ as well as $Y_1F=0$, $Y_2F=0$ where $Y_1:=D_m^{-1}\frac{\partial}{\partial a_{0,1}}D_m$, $Y_2:=D_m^{-1}\frac{\partial}{\partial b_{0,1}}D_m$.
Denote through $L_m$ the  Lie algebra generated by the operators
$X_1=\frac{\partial}{\partial a_{0,-1}}$,
$X_2=\frac{\partial}{\partial b_{0,-1}}$,
$Y_1$,
$Y_2$.
Algebra $L_m$ is called characteristic $m$-algebra.
Obviously,  operators $X_1$, $X_2$ are the f\/irst-order linear dif\/ferential operators, or vector f\/ields. The  operators $Y_1$, $Y_2$ can also be rewritten as vector f\/ields of the form
\begin{gather*}
Y_1=\frac{\partial}{\partial a_{0,0}} +\left(\frac{a_{1,0}}{a_{0,0}}-\frac{b_{1,0}b_{0,-1}}{a_{0,0}b_{0,0}a_{0,-1}}\right)\frac{\partial}{\partial a_{1,0}} +\left(\frac{a_{-1,0}}{a_{0,0}}+\frac{b_{0,-1}}{a_{0,0}a_{0,-1}}\right)\frac{\partial}{\partial a_{-1,0}} +\cdots \\
\hphantom{Y_1=}{}
 +\frac{1}{b_{0,-1}}\frac{\partial}{\partial b_{1,0}}
-\left(\frac{a_{-1,0}}{a_{0,0}b_{0,-1}}+\frac{1}{a_{0,0}a_{0,-1}}\right)\frac{\partial}{\partial b_{-1,0}}+\cdots, \\
Y_2=\frac{\partial}{\partial b_{0,0}}+ \left(\frac{b_{1,0}}{b_{0,0}}-\frac{a_{0,0}}{b_{0,0}b_{0,-1}}\right)\frac{\partial}{\partial b_{1,0}} +\left(\frac{b_{-1,0}}{b_{0,0}}+\frac{a_{-1,0}}{b_{0,0}b_{0,-1}}\right)\frac{\partial}{\partial b_{-1,0}}+\cdots.
\end{gather*}

We use the following lemma to show that $m$-algebra is of f\/inite dimension.

\begin{Lemma} \label{K=0_lemma}
Suppose that the vector field
\[
K=\sum_{k=1}^\infty \left( \alpha_k \frac{\partial}{\partial a_k} + \alpha_{-k} \frac{\partial}{\partial a_{-k}} \right) + \sum_{k=1}^\infty \left( \beta_k \frac{\partial}{\partial b_k} + \beta_{-k} \frac{\partial}{\partial b_{-k}} \right)
\]
satisfies the equality $D_nKD_n^{-1}=hK$, where $h$ is a function depending on shifts of variables~$a$ and~$b$, then $K=0$.
\end{Lemma}

Lemma can be proved by applying both sides of the equation $D_nKD_n^{-1}=hK$ to the dyna\-mi\-cal variables $a_{k}$, $b_{k}$.

One can easily check that
\begin{gather}
D_nX_1D_n^{-1}=\frac{a_{0,0}}{a_{1,0}}X_1,\qquad
D_nX_2D_n^{-1}=\frac{1}{a_{1,0}}X_1+\frac{b_{0,0}}{b_{1,0}}X_2,\nonumber\\
D_nY_1D_n^{-1}=\frac{a_{0,-1}}{a_{1,-1}}Y_1-\frac{a_{0,-1}}{a_{1,-1}b_{1,-1}}Y_2,\qquad
D_nY_2D_n^{-1}=\frac{b_{0,-1}}{b_{1,-1}}Y_2.\nonumber
\end{gather}

Put $\tilde{X}_1=a_{0,0}X_1$, $\tilde{X}_2=b_{0,0}X_2$, $\tilde{Y}_1=a_{0,-1} Y_1$, $\tilde{Y}_2=b_{0,-1} Y_2$, then
\begin{gather}
D_n\tilde{X}_1D_n^{-1}=\tilde{X}_1,\qquad
D_n\tilde{X}_2D_n^{-1}=\frac{b_{1,0}}{a_{0,0}a_{1,0}}\tilde{X}_1 +\tilde{X}_2,\nonumber\\
D_n\tilde{Y}_1D_n^{-1}=\tilde{Y}_1-\frac{a_{0,-1}}{b_{0,-1}b_{1,-1}}\tilde{Y}_2,\qquad
D_n\tilde{Y}_2D_n^{-1}=\tilde{Y}_2.\nonumber
\end{gather}

Taking commutators of the vector f\/ields $\tilde{X}_1$, $\tilde{X}_2$, $\tilde{Y}_1$, $\tilde{Y}_2$ we get vector f\/ields
\[
P_1=[X_1,Y_1], \qquad P_2=[X_2,Y_1],\qquad P_3=[X_2,Y_2].
\]

\begin{Lemma}\label{PP}
\begin{gather*}
D_nP_1D_n^{-1}=P_1-\frac{a_{0,0}}{b_{0,-1}b_{1,-1}}\tilde{Y}_2, \\
D_nP_2D_n^{-1}=P_2+\frac{b_{1,0}}{a_{0,0}a_{1,0}}P_1-
\frac{a_{0,-1}}{b_{0,-1}b_{1,-1}}P_3 \\
\hphantom{D_nP_2D_n^{-1}=}{}
+\left(\frac{a_{0,-1}b_{1,0}}{a_{0,0}^2a_{1,0}}+
\frac{a_{1,-1}b_{1,0}}{a_{0,0}a_{1,0}^2}\right)\tilde{X}_1+
\left(\frac{a_{0,-1}b_{0,0}}{b_{0,-1}^2b_{1,-1}}+
\frac{a_{1,-1}a_{0,0}b_{1,0}}{a_{1,0}b_{0,-1}b_{1,-1}^2}\right)\tilde{Y}_2, \\
D_nP_3D_n^{-1}=P_3-\frac{b_{1,-1}}{a_{0,0}a_{1,0}}\tilde{X}_1.
\end{gather*}
\end{Lemma}

Lemma can be proved by direct calculations. It allows to derive the following table of commutators which shows that the characteristic Lie algebra $L_m$ is of dimension seven:
\begin{center}
\begin {tabular}{|c|c|c|c|c|c|c|c|}
\hline
&$\tilde{X}_1$&$\tilde{X}_2$&$\tilde{Y}_1$&$\tilde{Y}_2$&$P_1$&$P_2$&$P_3$\\
\hline
$\tilde{X}_1$& $0$&$0$&$P_1$&$0$&$0$&$R_1$&$0$\\
\hline
$\tilde{X}_2$& $0$&$0$&$P_2$&$P_3$&$R_1$&$R_2$&$-2\tilde{X}_2$\\
\hline
$\tilde{Y}_1$& $-P_1$&$-P_2$&$0$&$0$&$2\tilde{Y}_1$&$R_3$&$R_4$\\
\hline
$\tilde{Y}_2$& $0$&$-P_3$&$0$&$0$&$0$&$R_4$&$0$\\
\hline
$P_1$&$0$&$-R_1$&$-2\tilde{Y}_1$&$0$&$0$&$Q_1$&$Q_2$\\
\hline
$P_2$&$-R_1$&$-R_2$&$-R_3$&$-R_4$&$-Q_1$&$0$&$Q_3$\\
\hline
$P_3$&$0$&$2\tilde{X}_2$&$-R_4$&$0$&$-Q_2$&$-Q_3$&$0$\\
\hline
\end{tabular}
\end{center}
Here the following notations are used
\begin{gather*}
R_1=\tilde{X}_2-\frac{b_{1,0}}{b_{0,-1}}P_1+\frac{b_{0,0}^2}{b_{0,-1}^2}\tilde{Y}_2, \\
R_2=\frac{2a_{0,-1}b_{0,0}^2}{a_{0,0}b_{0,-1}^2}P_3-\frac{2a_{0,-1}b_{0,0}^3}{a_{0,0}b_{0,-1}^3}\tilde{Y}_2+
\frac{2a_{0,-1}b_{0,0}}{a_{0,0}b_{0,-1}}\tilde{X}_2, \\
R_3=\frac{2a_{0,-1}^2b_{0,0}}{a_{0,0}^2b_{0,-1}}P_1-\frac{2a_{0,-1}b_{0,0}}{a_{0,0}b_{0,-1}}\tilde{Y}_1+
\frac{2a_{0,-1}^3b_{0,0}}{a_{0,0}^3b_{0,-1}}\tilde{X}_1, \\
R_4=-\tilde{Y_1}-\frac{a_{0,-1}^2}{a_{0,0}^2}\tilde{X}_1-\frac{a_{0,-1}}{a_{0,0}}P_1, \\
Q_1=\frac{a_{0,-1}b_{0,0}}{a_{0,0}b_{0,-1}}P_1-3P_2-P_3+
\frac{a_{0,-1}^2b_{0,0}}{a_{0,0}^2b_{0,-1}}\tilde{X}_1+\frac{b_{0,0}}{b_{0,-1}}\tilde{Y}_1, \\
Q_2=\frac{2b_{0,0}}{b_{0,-1}}\tilde{Y}_2-2P_3, \\
Q_3=-3P_2+\frac{a_{0,-1}b_{0,0}}{a_{0,0}b_{0,-1}}P_3-\frac{a_{0,-1}}{a_{0,0}}\tilde{X}_2-
\frac{a_{0,-1}b_{0,0}^2}{a_{0,0}b_{0,-1}^2}\tilde{Y}_2.
\end{gather*}
Due to the reasonings above any $m$-integral $F(\dots, a_{0,0},b_{0,0},a_{1,0},b_{1,0},a_{2,0},b_{2,0},\dots)$ should satisfy equations
\begin{gather}
\tilde{X}_1(F)=0,\qquad \tilde{X}_2(F)=0,\qquad\tilde{Y}_1(F)=0,\qquad \tilde{Y}_2(F)=0,\nonumber\\
P_1(F)=0,\qquad P_2(F)=0,\qquad P_3(F)=0.\label{sys_int}
\end{gather}

Solving system (\ref{sys_int}), it is enough to assume that $F$ depends on $b_{0,0}$, $a_{0,0}$, $b_{1,0}$, $a_{1,0}$, $b_{2,0}$, $a_{2,0}$ or, alternatively, $F$ depends on $a_{-1,0}$, $b_{0,0}$, $a_{0,0}$, $b_{1,0}$, $a_{1,0}$, $b_{2,0}$. Under such assumptions sys\-tem~(\ref{sys_int}) generates two systems of the f\/irst-order linear partial dif\/ferential equations which can be solved by Jacobi method. By solving these systems we f\/ind two independent $m$-integrals
\begin{gather*}
F_{(1)}=\frac{b_{0,0}}{b_{1,0}}+\frac{a_{0,0}b_{2,0}}{a_{1,0}b_{1,0}}+\frac{a_{2,0}}{a_{1,0}},\qquad
F_{(2)}=\frac{a_{-1,0}}{a_{0,0}}+\frac{a_{1,0}b_{0,0}}{a_{0,0}b_{1,0}}+\frac{b_{2,0}}{b_{1,0}}.
\end{gather*}

\section[Cutting off constraints for the Hirota equation and discrete Zakharov-Shabat systems]{Cutting of\/f constraints for the Hirota equation\\ and discrete Zakharov--Shabat systems}\label{section5}

In this section we construct Lax pairs for discrete systems corresponding to Cartan matrices of series $A_N$, $B_N$, $C_N$, $D_N^{(2)}$ and $A_1^{(1)}$. To this end we impose cutting of\/f constrains for the Hirota equation compatible with its Lax pair.

It is well known that Hirota chain
\begin{gather}
t^j_{0,0}t^j_{1,1}-t^j_{1,0}t^j_{0,1}=t^{j-1}_{1,0}t^{j+1}_{0,1} \label{eqHir}
\end{gather}
admits the Lax pair consisting  of two linear discrete equations \cite{DJM}
\begin{gather}\label{lax1}
\psi^j_{1,0}=\frac{t^{j+1}_{1,0}t^j_{0,0}}{t^{j+1}_{0,0}t^j_{1,0}}\psi^j_{0,0}-\psi^{j+1}_{0,0}, \qquad
\psi^j_{0,1}=\psi^{j}_{0,0}+\frac{t^{j+1}_{0,1}t^{j-1}_{0,0}}{t^j_{0,0}t^j_{0,1}}\psi^{j-1}_{0,0}.
\end{gather}
Here the lower indices mean as previously shifts of the arguments, and the upper index enume\-ra\-tes the f\/ield variables~$t^j$ and eigenfunctions~$\psi^j$.
Exclude from the system of equations~(\ref{lax1}) all the eigenfunctions with the upper index dif\/ferent from~$j$. As a result one gets a linear discrete hyperbolic equation for~$\psi^j$
\begin{gather}\label{eqn1.1}
\psi^j_{1,1}-\psi^j_{1,0}-\frac{t^{j+1}_{1,1}t^j_{0,1}}{t^{j+1}_{0,1}t^j_{1,1}}\psi^j_{0,1}+
\frac{t^j_{0,0}t^{j+1}_{1,1}}{t^j_{1,0}t^{j+1}_{0,1}}\psi^j_{0,0}=0.
\end{gather}
It is remarkable that by construction formulas (\ref{lax1}) def\/ine Laplace transformations for the linear hyperbolic equation~(\ref{eqn1.1}). Evidently, equation (\ref{eqHir}) is invariant under the  transformation def\/ined as $n\to 1-m$, $m\to 1-n$.
Under this transformation the Lax pair~(\ref{lax1})  transforms to a Lax pair
\begin{gather}\label{lax2}
y^j_{-1,0}=y^j_{0,0}+\frac{t^{j+1}_{-1,0}t^{j-1}_{0,0}}{t^j_{-1,0}t^j_{0,0}}y^{j-1}_{0,0}, \qquad
y^j_{0,-1}=\frac{t^{j+1}_{0,-1}t^j_{0,0}}{t^{j+1}_{0,0}t^j_{0,-1}}y^j_{0,0}-y^{j+1}_{0,0},
\end{gather} and equation (\ref{eqn1.1}) transforms to an equation
\begin{gather}\label{eqn2.1}
y^j_{1,1}-\frac{t^j_{1,0}t^{j+1}_{0,1}}{t^{j+1}_{0,0}t^j_{1,1}}y^j_{1,0}-\frac{t^j_{1,0}t^{j}_{0,1}}{t^{j}_{0,0}t^j_{1,1}}y^j_{0,1}
+\frac{t^j_{1,0}t^{j+1}_{0,1}}{t^{j+1}_{0,0}t^j_{1,1}}y^j_{0,0}=0.
\end{gather}

Put $y^j_{0,0}=\frac{t^{j+1}_{-1,0}}{t^j_{0,0}}g^j_{0,0}$, then Lax pair (\ref{lax2}) transforms to a Lax pair
\begin{gather}\label{lax3}
g^{j}_{-1,0}=\frac{t^{j+1}_{-1,0}t^{j}_{-1,0}}{t^{j}_{0,0}t^{j+1}_{-2,0}}
\big(g^{j}_{0,0}+g^{j-1}_{0,0}\big),\qquad
g^{j}_{0,-1}=\frac{t^{j+1}_{0,-1}t^{j+1}_{-1,0}}{t^{j+1}_{0,0}t^{j+1}_{-1,-1}}g^{j}_{0,0}-
\frac{t^{j}_{0,-1}t^{j+2}_{-1,0}}{t^{j+1}_{0,0}t^{j+1}_{-1,-1}}g^{j+1}_{0,0},
\end{gather} and equation (\ref{eqn2.1}) transforms to an equation
\begin{gather}\label{eqn3.1}
g^j_{1,1}-g^j_{1,0}-\frac{t^j_{1,0}t^{j+1}_{-1,1}}{t^{j+1}_{0,1}t^j_{0,0}}g^j_{0,1}+
\frac{t^j_{1,0}t^{j+1}_{-1,0}}{t^{j+1}_{0,0}t^j_{0,0}}g^j_{0,0}=0.
\end{gather}

Equations (\ref{lax3}) def\/ine Laplace transformations for the linear hyperbolic equation (\ref{eqn3.1}).
Thus we have two dif\/ferent Lax pairs for the Hirota chain and consequently two families (\ref{eqn1.1}), (\ref{eqn3.1}) of linear discrete hyperbolic equations enumerated by $j$. Study the question when  an equation from one family can be related, by a linear transformation, to  an equation from the other family. To this end we evaluate the Laplace invariants of these equations.

Recall that the Laplace invariants of a discrete hyperbolic type equation of the form
\begin{gather*}
a_{0,0}f_{1,1}+b_{0,0}f_{1,0}+c_{0,0}f_{0,1}+d_{0,0}f_{0,0}=0
\end{gather*}
are given by  (see~\cite{AdlerStartsev, D97, N02, Novikov})
\begin{gather*}
K_1=\frac{b_{0,0}c_{1,0}}{a_{0,0}d_{1,0}},\qquad
K_2=\frac{b_{0,1}c_{0,0}}{a_{0,0}d_{0,1}}.
\end{gather*}

By virtue of these formulas the invariants $K_{1\psi}$, $K_{2\psi}$ and $K_{1g}$, $K_{2g}$ of equations (\ref{eqn1.1}), (\ref{eqn3.1}) are, respectively,
\begin{gather*}
K_{1\psi}=\frac{t^j_{2,0}t^j_{1,1}}{t^j_{1,0}t^j_{2,1}},\qquad K_{2\psi}=\frac{t^{j+1}_{1,1}t^{j+1}_{0,2}}{t^{j+1}_{0,1}t^{j+1}_{1,2}}, \qquad
K_{1g}=\frac{t^{j+1}_{0,1}t^{j+1}_{1,0}}{t^{j+1}_{0,0}t^{j+1}_{1,1}},\qquad
K_{2g}=\frac{t^j_{1,0}t^j_{0,1}}{t^j_{0,0}t^j_{1,1}}.
\end{gather*}

It is known that two linear hyperbolic type equations are related to one another by a linear transformation only if their  corresponding Laplace invariants are equal. Evidently in generic case coincidence of the Laplace invariants generates two constraints on the f\/ield variables $t^j=t^j(n,m)$. Only for some special cases it gives only one constraint. We are interested in such special cases. For instance pair of equations $K_{1\psi}(n,m,j)=K_{1g}(n+1,m,j-1)$, $K_{2\psi}(n,m,j)=K_{2g}(n+1,m,j-1)$ is equivalent to the constraint
\begin{gather*}
t^{j-1}_{1,0}=t^{j+1}_{0,1}
\end{gather*}
which is interpreted as a cutting of\/f boundary condition for the chain (\ref{eqHir}).
For simplicity we put $j=0$, so the boundary condition becomes
\begin{gather}\label{bound_cond}
t^{-1}_{1,0}=t^1_{0,1}.
\end{gather}

\begin{Lemma}\label{lem_obr} Hirota equation \eqref{eqHir} is compatible with the reduction of the type of parity
\begin{gather*}
t^{-m}_{m+i,k}=t^{m}_{i,m+k}	
\end{gather*}
and boundary condition \eqref{bound_cond} is a consequence of this reduction.
\end{Lemma}

Following \cite{H} we can construct a Lax pair for the reduced chain.
Under the boundary condition (\ref{bound_cond}) we have coincidence of the invariants
\begin{gather*}
K_{1\psi}(n,m,j)=K_{1g}(n+1,m,j-1),\qquad K_{2\psi}(n,m,j)=K_{2g}(n+1,m,j-1),
\end{gather*}
and
\begin{gather*}
K_{1g}(n,m,j)=K_{1\psi}(n,m-1,j-1),\qquad
K_{2g}(n,m,j)=K_{2\psi}(n,m-1,j-1),
\end{gather*}
and  we can relate the eigenfunctions
\begin{gather*}
g^1_{0,1}=\lambda\psi^0_{0,0}, \qquad g^0_{1,0}=\lambda\psi^1_{0,0}.
\end{gather*}

We study the f\/inite reductions of the chain (\ref{eqHir}) on a f\/inite interval $N_L\leq j \leq N_R$. The reduction is obtained by imposing the boundary conditions at the left end-point $j=N_L$
\begin{gather}\label{L}
t^{N_L-1}_{1,0}=t^{N_L+1}_{0,1}
\end{gather}
and respectively at the right end-point $j=N_R$
\begin{gather}\label{R}
t^{N_R+1}_{0,1}=t^{N_R-1}_{1,0}.
\end{gather}

First we concentrate on the left end-point. Due to the reasonings above the eigenfunctions should satisfy the following gluing conditions
\begin{gather*}
\psi^{N_L-1}_{0,0}=\frac{1}{\lambda}g_{0,1}^{N_L}, \qquad g^{N_L-1}_{0,0}=\lambda\psi^{N_L}_{-1,0}.
\end{gather*}

These conditions allow one to close the Lax equations at the left end-point
\begin{gather}\label{dynam_g1}
g^{N_L}_{-1,0}=\frac{t^{N_L+1}_{-1,0}t^{N_L}_{-1,0}}{t^{N_L}_{0,0}t^{N_L+1}_{-2,0}}
\big(g^{N_L}_{0,0}+\lambda\psi^{N_L}_{-1,0}\big),\qquad
\psi^{N_L}_{0,1}=\psi^{N_L}_{0,0}+\frac{1}{\lambda}\frac{t^{N_L+1}_{0,1}t^{N_L+1}_{-1,1}}{t^{N_L}_{0,0}t^{N_L}_{0,1}}g^{N_L}_{0,1}.
\end{gather}

From  (\ref{lax1}), (\ref{lax3}) we have
\begin{gather*}
g^{N_L}_{0,-1}=\frac{t^{N_L+1}_{0,-1}t^{N_L+1}_{-1,0}}{t^{N_L+1}_{0,0}t^{N_L+1}_{-1,-1}}g^{N_L}_{0,0}-
\frac{t^{N_L}_{0,-1}t^{N_L+2}_{-1,0}}{t^{N_L+1}_{0,0}t^{N_L+1}_{-1,-1}}g^{N_L+1}_{0,0},\qquad
\psi^{N_L}_{1,0}=\frac{t^{N_L+1}_{1,0}t^{N_L}_{0,0}}{t^{N_L+1}_{0,0}t^{N_L}_{1,0}}\psi^{N_L}_{0,0}-\psi^{N_L+1}_{0,0}.
\end{gather*}

To derive similar equations at the point $N_R$ we use the right end-point constraint (\ref{R}), for which we have
\begin{gather*}
g^{N_R}_{0,0}=\psi^{N_R-1}_{0,-1}, \qquad \psi^{N_R}_{0,0}=g^{N_R-1}_{1,0}.
\end{gather*}

These conditions allow one to close the Lax equations at the right end-point
\begin{gather}
\psi^{N_R-1}_{1,0}=\frac{t^{N_R-1}_{0,0}t^{N_R}_{1,0}}{t^{N_R-1}_{1,0}t^{N_R}_{0,0}}\psi^{N_R-1}_{0,0}-g^{N_R-1}_{1,0},\nonumber\\
g^{N_R-1}_{0,-1}=\frac{t^{N_R}_{0,-1}t^{N_R}_{-1,0}}{t^{N_R}_{0,0}t^{N_R}_{-1,-1}}g^{N_R-1}_{0,0}-
\frac{\big(t^{N_R-1}_{0,-1}\big)^2}{t^{N_R}_{0,0}t^{N_R}_{-1,-1}}\psi^{N_R-1}_{0,-1}.\label{dynam_g2}
\end{gather}

From  (\ref{lax1}), (\ref{lax3}) we have
\begin{gather*}
g^{N_R-1}_{-1,0}=\frac{t^{N_R}_{-1,0}t^{N_R-1}_{-1,0}}{t^{N_R-1}_{0,0}t^{N_R}_{-2,0}}
\big(g^{N_R-1}_{0,0}+g^{N_R-2}_{0,0}\big),\qquad
\psi^{N_R-1}_{0,1}=\psi^{N_R-1}_{0,0}+\frac{t^{N_R}_{0,1}t^{N_R-2}_{0,0}}{t^{N_R-1}_{0,0}t^{N_R-1}_{0,1}}\psi^{N_R-2}_{0,0}.
\end{gather*}

The Lax pair found above is not convenient to work with because the operators contain shifts in opposite directions. Below we show that it can be rewritten in a usual form.

Shift of equation (\ref{dynam_g1}) forward with respect to the variable $n$ brings it to the form
\begin{gather*}
g^{N_L}_{1,0}=\frac{t^{N_L+1}_{-1,0}t^{N_L}_{1,0}}{t^{N_L}_{0,0}t^{N_L+1}_{0,0}}
g^{N_L}_{0,0}-\lambda\psi^{N_L}_{0,0},\\
g^{j}_{1,0}=\frac{t^{j+1}_{-1,0}t^j_{1,0}}{t^{j}_{0,0}t^{j+1}_{0,0}}
g^{j}_{0,0}-g^{j-1}_{1,0}=\sum\limits^{j}_{k=N_L}(-1)^{j-k}\frac{t^k_{1,0}t^{k+1}_{-1,0}}{t^k_{0,0}t^{k+1}_{0,0}}g^k_{0,0}+
(-1)^{j+1}\lambda\psi^{N_L}_{0,0},\\
 N_L+1 \leq j \leq N_R-1.
\end{gather*}

Shift of equation (\ref{dynam_g2}) forward with respect to the variable $m$ brings it to the form
\begin{gather*}
g^{N_R-1}_{0,1}=\frac{t^{N_R}_{0,1}t^{N_R}_{-1,0}}{t^{N_R}_{0,0}t^{N_R}_{-1,1}}g^{N_R-1}_{0,0}+
\frac{\big(t^{N_R-1}_{0,0}\big)^2}{t^{N_R}_{0,0}t^{N_R}_{-1,1}}\psi^{N_R-1}_{0,0},\\
g^{j}_{0,1}=\frac{t^{j+1}_{-1,0}t^{j+1}_{0,1}}{t^{j+1}_{0,0}t^{j+1}_{-1,1}}g^{j}_{0,0}+
\frac{t^{j+1}_{-1,0}t^{j+1}_{0,1}}{t^{j+1}_{0,0}t^{j+1}_{-1,1}}g^{j+1}_{0,1}=\\
\hphantom{g^{j}_{0,1}}{}
=\frac{t^{j+1}_{-1,0}t^{j+1}_{0,1}}{t^{j+1}_{0,0}t^{j+1}_{-1,1}}g^{j}_{0,0}
+\sum\limits^{N_R-1}_{k=j+1}\frac{t^j_{0,0}t^{k+1}_{-1,0}t^{k+1}_{0,1}}{t^{j+1}_{-1,1}t^{k}_{0,0}t^{k+1}_{0,0}}g^k_{0,0}+
\frac{t^j_{0,0}t^{N_R-1}_{0,0}}{t^{j+1}_{0,0}t^{N_R}_{0,0}}\psi^{N_R-1}_{0,0},
\qquad N_L \leq j \leq N_R-1.
\end{gather*}

So we have the following system of linear equations
\begin{gather}
\psi^{j}_{1,0}=\frac{t^{j+1}_{1,0}t^{j}_{0,0}}{t^{j+1}_{0,0}t^{j}_{1,0}}\psi^{j}_{0,0}-\psi^{j+1}_{0,0},\qquad N_L\leq j\leq N_R-2,\label{psi_NL}\\
\psi^{N_R-1}_{1,0}=\frac{t^{N_R-1}_{0,0}t^{N_R}_{1,0}}{t^{N_R-1}_{1,0}t^{N_R}_{0,0}}\psi^{N_R-1}_{0,0}
+\sum\limits^{N_R-1}_{k=N_L}(-1)^{N_R-k}\frac{t^k_{1,0}t^{k+1}_{-1,0}}{t^k_{0,0}t^{k+1}_{0,0}}g^k_{0,0}+
(-1)^{N_R-1}\lambda\psi^{N_L}_{0,0},\label{psi_N_Rn}\\
\psi^{N_L}_{0,1}=\psi^{N_L}_{0,0}+\frac{1}{\lambda}\frac{t^{N_L+1}_{0,1}t^{N_R-1}_{0,0}}{t^{N_L}_{0,1}t^{N_R}_{0,0}}\psi^{N_R-1}_{0,0}+
\sum\limits^{N_R-1}_{j=N_L}\frac{1}{\lambda}
\frac{t^{N_L+1}_{0,1}t^{k+1}_{0,1}t^{k+1}_{-1,0}}{t^{N_L}_{0,1}t^k_{0,0}t^{k+1}_{0,0}}g^k_{0,0},\label{psi_N_Lm}\\
\psi^{j}_{0,1}=\psi^{j}_{0,0}+\frac{t^{j}_{0,1}t^{j-2}_{0,0}}{t^{j-1}_{0,0}t^{j-1}_{0,1}}\psi^{j-1}_{0,0},\qquad N_L+1\leq j\leq N_R-1,\\
g^{j}_{1,0}=\sum\limits^{j}_{k=N_L}(-1)^{j-k}\frac{t^k_{1,0}t^{k+1}_{-1,0}}{t^k_{0,0}t^{k+1}_{0,0}}g^k_{0,0}+
(-1)^{j+1}\lambda\psi^{N_L}_{0,0}, \qquad N_L \leq j \leq N_R-1,\label{gLax_jn}\\
g^{j}_{0,1}=\frac{t^{j+1}_{-1,0}t^{j+1}_{0,1}}{t^{j+1}_{0,0}t^{j+1}_{-1,1}}g^{j}_{0,0}
+\sum\limits^{N_R-1}_{k=j+1}\frac{t^j_{0,0}t^{k+1}_{-1,0}t^{k+1}_{0,1}}{t^{j+1}_{-1,1}t^{k}_{0,0}t^{k+1}_{0,0}}g^k_{0,0}+
\frac{t^j_{0,0}t^{N_R-1}_{0,0}}{t^{j+1}_{0,0}t^{N_R}_{0,0}}\psi^{N_R-1}_{0,0},\nonumber\\  N_L \leq j \leq N_R-1.\label{gLax_jm}
\end{gather}

In the next subsection we gather these equations to a matrix form.

\subsection[Lax pair for systems corresponding to the algebras $D^{(2)}_N$, $A^{(1)}_1$]{Lax pair for systems corresponding to the algebras $\boldsymbol{D^{(2)}_N}$, $\boldsymbol{A^{(1)}_1}$}

Imposing of non-degenerate boundary conditions (\ref{L}), (\ref{R}) at both end-points $N_L=0$ and $N_R=N$ leads to the system corresponding to the matrix $D^{(2)}_{N+1}$, $N\geq 2$
\begin{gather}
t^0_{0,0}t^0_{1,1}-t^0_{1,0}t^0_{0,1}=\big(t^1_{0,1}\big)^2,\qquad
t^j_{0,0}t^j_{1,1}-t^j_{1,0}t^j_{0,1}=t^{j-1}_{1,0}t^{j+1}_{0,1},\qquad 1\leq j \leq N-1,\label{chain1_u}\\
t^N_{0,0}t^N_{1,1}-t^N_{1,0}t^N_{0,1}=\big(t^{N-1}_{1,0}\big)^2.\nonumber
\end{gather}
Write the set of equations (\ref{psi_NL})--(\ref{gLax_jm}) in the form of a Lax pair for system~(\ref{chain1_u}). Denote the eigenvector as follow $P=(\psi^0,\psi^1,\dots,\psi^{N-1}, g^0,g^1,\dots,g^{N-1})^T$. Introduce $2N \times 2N$ matrices
\begin{gather*}
A=\left(\!
\begin{array}{@{}c@{\,\,}c@{\,\,}c@{\,\,}c@{\,\,}c@{\,\,}c@{\,\,}c@{\,\,}c@{}}
\frac{t^1_{1,0}t^0_{0,0}}{t^1_{0,0}t^0_{1,0}}&-1&0&\dots&0&0&\dots&0\\
0&\frac{t^{2}_{1,0}t^1_{0,0}}{t^{2}_{0,0}t^1_{1,0}}&-1&\dots&0&0&\dots&0\\
&&&\dots\\
(-1)^{N-1}\lambda&0&\dots&\frac{t^{N}_{1,0}t^{N-1}_{0,0}}{t^{N}_{0,0}t^{N-1}_{1,0}}&
(-1)^{N}\frac{t^0_{1,0}t^1_{-1,0}}{t^0_{0,0}t^1_{0,0}}&(-1)^{N-1}\frac{t^1_{1,0}t^2_{-1,0}}{t^1_{0,0}t^2_{0,0}}&
\dots&-\frac{t^{N-1}_{1,0}t^{N}_{-1,0}}{t^{N-1}_{0,0}t^{N}_{0,0}}\\
-\lambda&0&\dots&0&\frac{t^0_{1,0}t^1_{-1,0}}{t^0_{0,0}t^1_{0,0}}&0&\dots&0\\
\lambda&0&\dots&0&-\frac{t^0_{1,0}t^1_{-1,0}}{t^0_{0,0}t^1_{0,0}}&\frac{t^1_{1,0}t^2_{-1,0}}{t^1_{0,0}t^2_{0,0}}&\dots&0\\
&&&\dots\\
(-1)^{i+1}\lambda&0&\dots&0&\dots&
(-1)^{i-j}\frac{t^j_{1,0}t^{j+1}_{-1,0}}{t^j_{0,0}t^{j+1}_{0,0}}&\dots&0\\
&&&\dots\\
(-1)^{N}\lambda&0&\dots&0&(-1)^{N-1}\frac{t^0_{1,0}t^1_{-1,0}}{t^0_{0,0}t^1_{0,0}}&
(-1)^{N}\frac{t^1_{1,0}t^2_{-1,0}}{t^1_{0,0}t^2_{0,0}}&\dots&\frac{t^{N-1}_{1,0}t^{N}_{-1,0}}{t^{N-1}_{0,0}t^{N}_{0,0}}
\end{array}\!\right),
\\
B=\left(\!
\begin{array}{@{}c@{\,\,}c@{\,\,}c@{\,\,}c@{\,\,}c@{\,\,}c@{\,\,}c@{\,\,}c@{}}
1&0&\dots&\frac{1}{\lambda}\frac{t^1_{0,1}t^{N-1}_{0,0}}{t^0_{0,1}t^N_{0,0}}&
\frac{1}{\lambda}\frac{(t^1_{0,1})^2t^1_{-1,0}}{t^0_{0,0}t^0_{0,1}t^1_{0,0}}&
\frac{1}{\lambda}\frac{t^1_{0,1}t^2_{-1,0}t^2_{0,1}}{t^0_{0,1}t^1_{0,0}t^2_{0,0}}&\dots&
\frac{1}{\lambda}\frac{t^1_{0,1}t^N_{-1,0}t^N_{0,1}}{t^0_{0,1}t^{N-1}_{0,0}t^N_{0,0}}\\
\frac{t^0_{0,0}t^2_{0,1}}{t^1_{0,0}t^1_{0,1}}&1&\dots&0&0&0&\dots&0\\
&&\dots&&\\
0&\dots&\frac{t^{N-2}_{0,0}t^N_{0,1}}{t^{N-1}_{0,0}t^{N-1}_{0,1}}&1&0&0&\dots&0\\
0&\dots&0&\frac{t^0_{0,0}t^{N-1}_{0,0}}{t^1_{-1,1}t^N_{0,0}}&\frac{t^1_{0,1}t^1_{-1,0}}{t^1_{0,0}t^1_{-1,1}}&
\frac{t^0_{0,0}t^2_{-1,0}t^2_{0,1}}{t^1_{-1,1}t^1_{0,0}t^2_{0,0}}&\dots&
\frac{t^0_{0,0}t^N_{-1,0}t^N_{0,1}}{t^1_{-1,1}t^{N-1}_{0,0}t^N_{0,0}}\\
0&\dots&0&\frac{t^1_{0,0}t^{N-1}_{0,0}}{t^2_{-1,1}t^N_{0,0}}&0&
\frac{t^2_{-1,0}t^2_{0,1}}{t^2_{0,0}t^2_{-1,1}}&\dots&
\frac{t^1_{0,0}t^N_{-1,0}t^N_{0,1}}{t^2_{-1,1}t^{N-1}_{0,0}t^N_{0,0}}\\
&\dots\\
0&\dots&0&\frac{(t^{N-1}_{0,0})^2}{t^N_{-1,1}t^N_{0,0}}&0&
0&\dots&
\frac{t^N_{-1,0}t^N_{0,1}}{t^{N}_{0,0}t^N_{-1,1}}
\end{array}\!\right).
\end{gather*}
It is straightforward to check that the compatibility condition of the equations
\begin{gather}\label{Pn10}
P_{1,0}=AP,\qquad P_{0,1}=BP
\end{gather}
is equivalent the system (\ref{chain1_u}).

\begin{Example}\label{ex1}
Consider a particular case when $N_L=0$ and $N_R=1$
\begin{gather*}
t^0_{0,0}t^0_{1,1}-t^0_{1,0}t^0_{0,1}=\big(t^1_{0,1}\big)^2,\qquad
t^1_{0,0}t^1_{1,1}-t^1_{1,0}t^1_{0,1}=\big(t^0_{1,0}\big)^2.
\end{gather*}

The system corresponds to algebra $A^{(1)}_1$. Its Lax pair is given by (\ref{Pn10}) with the matrices
\[
A=\left(
\begin{array}{@{}cc@{}}
\frac{t^1_{1,0}t^0_{0,0}}{t^1_{0,0}t^0_{1,0}}+\lambda&-\frac{t^0_{1,0}t^1_{-1,0}}{t^0_{0,0}t^1_{0,0}}\vspace{1mm}\\
-\lambda&\frac{t^0_{1,0}t^1_{-1,0}}{t^0_{0,0}t^1_{0,0}}
\end{array}\right),\qquad
B=\left(
\begin{array}{@{}cc@{}}
1+\frac{1}{\lambda}\frac{t^1_{0,1}t^0_{0,0}}{t^0_{0,1}t^1_{0,0}}&
\frac{1}{\lambda}\frac{(t^1_{0,1})^2t^1_{-1,0}}{t^0_{0,0}t^0_{0,1}t^1_{0,0}}\vspace{1mm}\\
\frac{(t^0_{0,0})^2}{t^1_{0,0}t^1_{-1,1}}&\frac{t^1_{0,1}t^1_{-1,0}}{t^1_{0,0}t^1_{-1,1}}
\end{array}\right).
\]
\end{Example}

\begin{Example}\label{ex2}
If one imposes condition (\ref{L}) for $N_L=0$ and (\ref{R}) for $N_R=2$, then
\begin{gather*}
t^0_{0,0}t^0_{1,1}-t^0_{1,0}t^0_{0,1}=\big(t^1_{0,1}\big)^2,\qquad
t^1_{0,0}t^1_{1,1}-t^1_{1,0}t^1_{0,1}=t^0_{1,0}t^2_{0,1},\qquad
t^2_{0,0}t^2_{1,1}-t^2_{1,0}t^2_{0,1}=\big(t^1_{1,0}\big)^2.
\end{gather*}
The Lax pair is of the form (\ref{Pn10}) where $A$ and $B$ are $4\times 4$ matrices
\begin{gather*}
A=\left(
\begin{array}{@{}cccc@{}}
\frac{t^1_{1,0}t^0_{0,0}}{t^1_{0,0}t^0_{1,0}}&-1&0&0\vspace{1mm}\\
-\lambda&\frac{t^{2}_{1,0}t^1_{0,0}}{t^{2}_{0,0}t^1_{1,0}}&
\frac{t^0_{1,0}t^1_{-1,0}}{t^0_{0,0}t^1_{0,0}}&-\frac{t^1_{1,0}t^2_{-1,0}}{t^1_{0,0}t^2_{0,0}}\vspace{1mm}\\
-\lambda&0&\frac{t^0_{1,0}t^1_{-1,0}}{t^0_{0,0}t^1_{0,0}}&0\vspace{1mm}\\
\lambda&0&-\frac{t^0_{1,0}t^1_{-1,0}}{t^0_{0,0}t^1_{0,0}}&\frac{t^1_{1,0}t^2_{-1,0}}{t^1_{0,0}t^2_{0,0}}
\end{array}\right),\\
B=\left(
\begin{array}{@{}cccc@{}}
1&\frac{1}{\lambda}\frac{t^1_{0,1}t^1_{0,0}}{t^0_{0,1}t^2_{0,0}}&
\frac{1}{\lambda}\frac{(t^1_{0,1})^2t^1_{-1,0}}{t^0_{0,0}t^0_{0,1}t^1_{0,0}}&
\frac{1}{\lambda}\frac{t^1_{0,1}t^2_{-1,0}t^2_{0,1}}{t^0_{0,1}t^1_{0,0}t^2_{0,0}}\vspace{1mm}\\
\frac{t^0_{0,0}t^2_{0,1}}{t^1_{0,0}t^1_{0,1}}&1&0&0\vspace{1mm}\\
0&\frac{t^0_{0,0}t^1_{0,0}}{t^1_{-1,1}t^2_{0,0}}&\frac{t^1_{0,1}t^1_{-1,0}}{t^1_{0,0}t^1_{-1,1}}&
\frac{t^0_{0,0}t^2_{-1,0}t^2_{0,1}}{t^1_{-1,1}t^1_{0,0}t^2_{0,0}}\vspace{1mm}\\
0&\frac{(t^1_{0,0})^2}{t^2_{-1,1}t^2_{0,0}}&0&
\frac{t^2_{-1,0}t^2_{0,1}}{t^2_{0,0}t^2_{-1,1}}
\end{array}\right).
\end{gather*}
\end{Example}

\begin{Remark}In Example \ref{ex1} we give a Lax pair realized in $2\times 2$ matrices, while general formula~\eqref{Pn10} generates $3\times 3$ matrices. The matter is that in the Lax pair obtained directly from~\eqref{Pn10} we made in this case some additional reduction.
\end{Remark}

\subsection[Lax pair for systems corresponding to the algebras  $A_N$]{Lax pair for systems corresponding to the algebras  $\boldsymbol{A_N}$}

Instead of boundary condition (\ref{bound_cond}) we can
use also the degenerate boundary conditions of the form
\begin{gather}\label{Ldeg_bound_cond}
t^{N_L-1}=0,\qquad t^{N_L}=1
\end{gather}
at the left end-point and
\begin{gather}\label{Rdeg_bound_cond}
t^{N_R+1}=0,\qquad t^{N_R}=1
\end{gather}
at the right end-point. The degenerate boundary conditions imply that the corresponding eigenfunctions are zero: $g^{N_L-1}=0$ and $\psi^{N_R}=0$.

In order to obtain the system corresponding to the Cartan matrix $A_{N-1}$
we cut of\/f the Hirota chain by imposing degenerate boundary conditions (\ref{Ldeg_bound_cond}) at the point $N_L=0$ and (\ref{Rdeg_bound_cond}) at $N_R=N$. The resulting reduction is as follows
\begin{gather}
t^1_{0,0}t^1_{1,1}-t^1_{1,0}t^1_{0,1}=t^2_{0,1},\qquad
t^j_{0,0}t^j_{1,1}-t^j_{1,0}t^j_{0,1}=t^{j-1}_{1,0}t^{j+1}_{0,1},\qquad 2\leq j \leq N-2,\label{chain_deg}\\
t^{N-1}_{0,0}t^{N-1}_{1,1}-t^{N-1}_{1,0}t^{N-1}_{0,1}=t^{N-2}_{1,0}.\nonumber
\end{gather}
In this case our algorithm gives the Lax pair found years ago in \cite{Hirota}. In order to formulate it introduce the eigenvector $P=(\psi^0,\psi^1,\dots,\psi^{N-1})^T$ and $N\times N$ matrices
 \begin{gather*}
U=\left(
\begin{array}{@{}c@{\,\,}c@{\,\,}c@{\,\,}c@{\,\,}c@{}}
\frac{t^1_{1,0}}{t^1_{0,0}}&-1&\dots&0&0 \\
0&\frac{t^{2}_{1,0}t^1_{0,0}}{t^{2}_{0,0}t^1_{1,0}}&\dots&0&0 \\
&&\dots\\
0&0&\dots&\frac{t^{N-1}_{1,0}t^{N-2}_{0,0}}{t^{N-1}_{0,0}t^{N-2}_{1,0}}&-1 \\
0&0&\dots&0&\frac{t^{N-1}_{0,0}}{t^{N-1}_{1,0}}
\end{array}\right),\qquad
V=\left(
\begin{array}{@{}c@{\,\,}c@{\,\,}c@{\,\,}c@{\,\,}c@{\,\,}c@{}}
1&0&0&\dots&0&0\\
\frac{t^2_{0,1}}{t^1_{0,0}t^1_{0,1}}&1&0&\dots&0&0 \\
0&\frac{t^{3}_{0,1}t^{1}_{0,0}}{t^2_{0,0}t^2_{0,1}}&1&\dots&0&0 \\
&&\dots\\
0&0&\dots&0&\frac{t^{N-2}_{0,0}}{t^{N-1}_{0,0}t^{N-1}_{0,1}}&1
\end{array}\right).
 \end{gather*}
It is easy to check that the compatibility condition of the equations
\begin{gather*}
P_{1,0}=UP,\qquad P_{0,1}=VP.
\end{gather*}
leads to the system (\ref{chain_deg}).

\subsection[Lax pair for systems corresponding to the algebras  $B_N$]{Lax pair for systems corresponding to the algebras  $\boldsymbol{B_N}$}

We impose boundary condition (\ref{Ldeg_bound_cond}) for $N_L=0$ and (\ref{R}) for $N_R=N$. The resulting reduction is as follows
\begin{gather}
t^1_{0,0}t^1_{1,1}-t^1_{1,0}t^1_{0,1}=t^2_{0,1},\qquad
t^j_{0,0}t^j_{1,1}-t^j_{1,0}t^j_{0,1}=t^{j-1}_{1,0}t^{j+1}_{0,1},\qquad 2\leq j \leq N-1,\label{chain4_u}\\
t^{N}_{0,0}t^{N}_{1,1}-t^{N}_{1,0}t^{N}_{0,1}=\big(t^{N-1}_{1,0}\big)^2.\nonumber
\end{gather}
System (\ref{chain4_u}) can be rewritten in form of  (\ref{dis-dis-system}) by changing the variables $t^j=e^{-u^j}$. It corresponds to the algebra $B_N$. The system admits a Lax pair.
Let us denote $P=(\psi^0,\psi^1,\dots,\psi^{N-1}$, $g^0,g^1, \dots,g^{N-1})^T$ and introduce $2N\times 2N$ matrices
\begin{gather}
A=\left(
\begin{array}{@{}c@{\,\,}c@{\,\,}c@{\,\,}c@{\,\,}c@{\,\,}c@{\,\,}c@{\,\,}c@{}}
\frac{t^1_{1,0}}{t^1_{0,0}}&-1&0&\dots&0&0&\dots&0\\
0&\frac{t^{2}_{1,0}t^1_{0,0}}{t^{2}_{0,0}t^1_{1,0}}&-1&\dots&0&0&\dots&0\\
&&&\dots\\
0&0&\dots&\frac{t^{N}_{1,0}t^{N-1}_{0,0}}{t^{N}_{0,0}t^{N-1}_{1,0}}&
(-1)^{N}\frac{t^1_{-1,0}}{t^1_{0,0}}&(-1)^{N-1}\frac{t^1_{1,0}t^2_{-1,0}}{t^1_{0,0}t^2_{0,0}}&
\dots&-\frac{t^{N-1}_{1,0}t^{N}_{-1,0}}{t^{N-1}_{0,0}t^{N}_{0,0}}\\
0&0&\dots&0&\frac{t^1_{-1,0}}{t^1_{0,0}}&0&\dots&0\\
0&0&\dots&0&-\frac{t^1_{-1,0}}{t^1_{0,0}}&\frac{t^1_{1,0}t^2_{-1,0}}{t^1_{0,0}t^2_{0,0}}&\dots&0\\
&&&\dots\\
0&0&\dots&0&(-1)^{N-1}\frac{t^1_{-1,0}}{t^1_{0,0}}&
(-1)^{N}\frac{t^1_{1,0}t^2_{-1,0}}{t^1_{0,0}t^2_{0,0}}&\dots&\frac{t^{N-1}_{1,0}t^{N}_{-1,0}}{t^{N-1}_{0,0}t^{N}_{0,0}}
\end{array}\right),\label{A_for_B_N}\\
B=\left(
\begin{array}{@{}c@{\,\,}c@{\,\,}c@{\,\,}c@{\,\,}c@{\,\,}c@{\,\,}c@{\,\,}c@{}}
1&0&\dots&0&0&0&\dots&0\\
\frac{t^2_{0,1}}{t^1_{0,0}t^1_{0,1}}&1&\dots&0&0&0&\dots&0\\
0&\frac{t^1_{0,0}t^3_{0,1}}{t^2_{0,0}t^2_{0,1}}&\dots&0&0&0&\dots&0\\
&&\dots&&\\
0&\dots&\frac{t^{N-2}_{0,0}t^N_{0,1}}{t^{N-1}_{0,0}t^{N-1}_{0,1}}&1&0&0&\dots&0\\
0&\dots&0&\frac{t^{N-1}_{0,0}}{t^1_{-1,1}t^N_{0,0}}&\frac{t^1_{0,1}t^1_{-1,0}}{t^1_{0,0}t^1_{-1,1}}&
\frac{t^2_{-1,0}t^2_{0,1}}{t^1_{-1,1}t^1_{0,0}t^2_{0,0}}&\dots&
\frac{t^N_{-1,0}t^N_{0,1}}{t^1_{-1,1}t^{N-1}_{0,0}t^N_{0,0}}\\
0&\dots&0&\frac{t^1_{0,0}t^{N-1}_{0,0}}{t^2_{-1,1}t^N_{0,0}}&0&
\frac{t^2_{-1,0}t^2_{0,1}}{t^2_{0,0}t^2_{-1,1}}&\dots&
\frac{t^1_{0,0}t^N_{-1,0}t^N_{0,1}}{t^2_{-1,1}t^{N-1}_{0,0}t^N_{0,0}}\\
&\dots\\
0&\dots&0&\frac{(t^{N-1}_{0,0})^2}{t^N_{-1,1}t^N_{0,0}}&0&
0&\dots&
\frac{t^N_{-1,0}t^N_{0,1}}{t^{N}_{0,0}t^N_{-1,1}}
\end{array}\right).\label{B_for_B_N}
\end{gather}
Then according to our general scheme the compatibility condition of the equations
\begin{gather*}
P_{1,0}=AP,\qquad P_{0,1}=BP
\end{gather*}
leads to the system (\ref{chain4_u}).

\begin{Remark} The system $B_N$ can be obtained from the system $A_{2N-1}$ by imposing the cutting of\/f constraint of the form
\begin{gather*}
t^{N+1}_{0,1}=t^{N-1}_{1,0}
\end{gather*}
(see Lemma~\ref{lem_obr} above).
\end{Remark}

\section[Method of finding integrals from Lax representation for the systems corresponding to the Cartan matrices $A_N$, $B_N$]{Method of f\/inding integrals from Lax representation for\\ the systems corresponding to the Cartan matrices $\boldsymbol{A_N}$, $\boldsymbol{B_N}$}\label{section6}

In this section we show that the Lax pair allows one to generate integrals for the systems corresponding to the simple Lie algebras $A_N$, $B_N$. Concentrate on $m$-integrals. Due to the def\/inition we have two dif\/ferent expression for the shifted eigenvector $P_{k,1}$, $k\geq 1$
\[P_{k,1}=V_{k,0}U_{k-1,0}U_{k-2,0}\cdots U_{1,0}U_{0,0}P,\]
and similarly
\[P_{k,1}=U_{k-1,1}U_{k-2,1}\cdots U_{1,1}U_{0,1}V_{0,0}P.\]
Comparison of these two formulas yields
\begin{gather*}
D_m(U_{k-1,0}U_{k-2,0}\cdots U_{1,0}U_{0,0})=V_{k,0}U_{k-1,0}U_{k-2,0}\cdots U_{1,0}U_{0,0}V^{-1}_{0,0}.
\end{gather*}
Due to the triangular structure of the matrices $V_{k,0}$ and $V_{0,0}$ the map converting any upper triangular matrix $X$ to a matrix $\bar X=V_{k,0}XV^{-1}_{0,0}$ leaves unchanged the element of the mat\-rix~$X$ located at the right upper corner: $(X)_{1,N}=\bar X_{1,N}$. Thus the corresponding element, denote it through  $I_{(k-N)}$ ($k>N$), of the upper triangular matrix $U_{k-1,0}U_{k-2,0}\cdots U_{1,0}U_{0,0}$ is an $m$-integral. In such a way we get a set of integrals $I_{(1)}, I_{(2)}, \dots , I_{(N)}$. Examples below show that they constitute a complete set of $m$-integrals, however we are not able to prove this fact in general.  In a similar way one can f\/ind integrals in the other direction.

Let us illustrate the statement above with the following
\begin{Example}\label{counterex}
Consider the system (\ref{chain_deg}) for $N=3$
\begin{gather*}
t^1_{0,0}t^1_{1,1}-t^1_{1,0}t^1_{0,1}=t^2_{0,1},\qquad
t^2_{0,0}t^2_{1,1}-t^2_{1,0}t^2_{0,1}=t^1_{1,0}.
\end{gather*}
Recall its  Lax pair
\begin{gather*}
P_{1,0}=UP,\qquad P_{0,1}=VP,
\end{gather*}
where $P=(\psi^0,\psi^1,\psi^{2})^T$ and
\begin{gather*}
U=
\begin{pmatrix}
\frac{t^1_{1,0}}{t^1_{0,0}}&-1&0\\
0&\frac{t^{2}_{1,0}t^1_{0,0}}{t^{2}_{0,0}t^1_{1,0}}&-1\\
0&0&\frac{t^2_{0,0}}{t^2_{1,0}}
\end{pmatrix},\qquad
V=
\begin{pmatrix}
1&0&0\\
\frac{t^2_{0,1}}{t^1_{0,0}t^1_{0,1}}&1&0\\
0&\frac{t^{1}_{0,0}}{t^2_{0,0}t^2_{0,1}}&1
\end{pmatrix}.
\end{gather*}
Evaluate the elements at the right upper corner for the following two products $U_{2,0}U_{1,0}U_{0,0}$ and $U_{3,0}U_{2,0}U_{1,0}U_{0,0}$ and f\/ind two independent $m$-integrals
\[
I_{(1)}=\frac{t^1_3}{t^1_2}+\frac{t^1_1t^2_2}{t^1_2t^2_1}+\frac{t^2_0}{t^2_1},\qquad
I_{(2)}=\frac{t^1_4}{t^1_3}I_{(1)}+\frac{t^1_1t^2_3}{t^1_3t^2_1}+
\frac{t^1_2t^2_0t^2_3}{t^1_3t^2_1t^2_2}+\frac{t^2_0}{t^2_2},
\]
where the second index for the variables $t^1$, $t^2$ is omitted.
Since the integral~$I_{(2)}$ is too complicated one can choose a more simple one
\[
\tilde I_{(2)}=D_n^{-1}\big(I_{(1)}D_nI_{(1)}-I_{(2)}\big)=\frac{t^1_{0}}{t^1_{1}}+\frac{t^1_{2}t^2_0}{t^1_{1}t^2_1}
+\frac{t^2_2}{t^2_1}.
\]
By using  Theorem~\ref{independent_int} one can check that integrals~$I_{(1)}$ and~$\widetilde I_{(2)}$ provide a complete set of integrals.
\end{Example}

In a similar way integrals for the system (\ref{chain4_u}), corresponding to the algebra $B_N$, are constructed.
Let us consider a matrix~$\Phi$ of dimension $2N\times 2N$
\begin{gather}\label{Phi}
\Phi=
\begin{pmatrix}
E_{11}&E_{12}\\
E_{21}&E_{22}
\end{pmatrix},
\end{gather}
where $E_{11}$ is the unity matrix of dimension $N$, $E_{12}$, $E_{21}$ are matrices of dimension $N$ with zero entries, $E_{22}$ is a matrix of dimension $N$ with the unity adverse diagonal.
By an automorphism $X\rightarrow\Phi X\Phi^{-1}$ the matrices (\ref{A_for_B_N}) and (\ref{B_for_B_N}) are transformed to the triangular matrices
\begin{gather}\nonumber
\bar{A}=\left(
\begin{array}{@{}c@{\,\,}c@{\,\,}c@{\,\,}c@{\,\,}c@{\,\,}c@{\,\,}c@{\,\,}c@{}}
\frac{t^1_{1,0}}{t^1_{0,0}}&-1&0&\dots&0&0&\dots&0\\
0&\frac{t^{2}_{1,0}t^1_{0,0}}{t^{2}_{0,0}t^1_{1,0}}&-1&\dots&0&0&\dots&0\\
&&&\dots\\
0&0&\dots&\frac{t^{N}_{1,0}t^{N-1}_{0,0}}{t^{N}_{0,0}t^{N-1}_{1,0}}&
-\frac{t^{N-1}_{1,0}t^{N}_{-1,0}}{t^{N-1}_{0,0}t^{N}_{0,0}}
&\frac{t^{N-2}_{1,0}t^{N-1}_{-1,0}}{t^{N-2}_{0,0}t^{N-1}_{0,0}}&
\dots&(-1)^{N}\frac{t^1_{-1,0}}{t^1_{0,0}}\\
0&0&\dots&0&\frac{t^{N-1}_{1,0}t^{N}_{-1,0}}{t^{N-1}_{0,0}t^{N}_{0,0}}&
-\frac{t^{N-2}_{1,0}t^{N-1}_{-1,0}}{t^{N-2}_{0,0}t^{N-1}_{0,0}}&\dots&(-1)^{N-1}\frac{t^1_{-1,0}}{t^1_{0,0}}\\
0&0&\dots&0&0&\frac{t^{N-2}_{1,0}t^{N-1}_{-1,0}}{t^{N-2}_{0,0}t^{N-1}_{0,0}}&\dots&(-1)^{N}\frac{t^1_{-1,0}}{t^1_{0,0}}\\
&&&\dots\\
0&0&\dots&0&0&0&\dots&\frac{t^1_{-1,0}}{t^1_{0,0}}
\end{array}\right),
\\
\label{bar_B}
\bar{B}=\left(
\begin{array}{@{}c@{\,\,}c@{\,\,}c@{\,\,}c@{\,\,}c@{\,\,}c@{\,\,}c@{\,\,}c@{}}
1&0&\dots&0&0&0&\dots&0\\
\frac{t^2_{0,1}}{t^1_{0,0}t^1_{0,1}}&1&\dots&0&0&0&\dots&0\\
0&\frac{t^1_{0,0}t^3_{0,1}}{t^2_{0,0}t^2_{0,1}}&\dots&0&0&0&\dots&0\\
&&\dots&&\\
0&\dots&\frac{t^{N-2}_{0,0}t^N_{0,1}}{t^{N-1}_{0,0}t^{N-1}_{0,1}}&1&0&0&\dots&0\\
0&\dots&0&\frac{t^{N-1}_{0,0}}{t^1_{-1,1}t^N_{0,0}}&\frac{t^N_{-1,0}t^N_{0,1}}{t^{N}_{0,0}t^N_{-1,1}}&
0&\dots&0\\
0&\dots&0&\frac{t^1_{0,0}t^{N-1}_{0,0}}{t^2_{-1,1}t^N_{0,0}}&
\frac{t^{N-2}_{0,0}t^N_{-1,0}t^N_{0,1}}{t^{N-1}_{-1,1}t^{N-1}_{0,0}t^N_{0,0}}&
\frac{t^{N-1}_{-1,0}t^{N-1}_{0,1}}{t^{N-1}_{0,0}t^{N-1}_{-1,1}}&\dots&0\\
&\dots\\
0&\dots&0&\frac{(t^{N-1}_{0,0})^2}{t^N_{-1,1}t^N_{0,0}}&\frac{t^N_{-1,0}t^N_{0,1}}{t^1_{-1,1}t^{N-1}_{0,0}t^N_{0,0}}
&\frac{t^{N-1}_{-1,0}t^{N-1}_{0,1}}{t^1_{-1,1}t^{N-2}_{0,0}t^{N-1}_{0,0}}&\dots&\frac{t^1_{0,1}t^1_{-1,0}}{t^1_{0,0}t^1_{-1,1}}
\end{array}\right).
\end{gather}

By an automorphism acting as follows $X\rightarrow F^{-1}XF$  matrix (\ref{bar_B}) is transformed to a lower triangular matrix for which all diagonal entries are equal to the unity. Here~$F$ is a $2N\times 2N$ matrix of the form
\[
F=
\begin{pmatrix}
E_{11}&E_{12}\\
E_{21}&F_{22}
\end{pmatrix},
\]
$F_{22}$ is a diagonal matrix such that $F_{22}=\operatorname{diag}\left(\frac{t^N_{0,0}}{t^N_{-1,0}},\frac{t^{N-1}_{0,0}}{t^{N-1}_{-1,0}},\dots, \frac{t^1_{0,0}}{t^1_{-1,0}}\right)$ and $E_{i,j}$ are def\/ined in~(\ref{Phi}). Reasonings similar to that of the case $A_N$ allow one to derive the integrals.

\begin{Example}
Consider the system (\ref{chain4_u}) for $N=2$
\begin{gather*}
t^1_{0,0}t^1_{1,1}-t^1_{1,0}t^1_{0,1}=t^2_{0,1},\qquad
t^2_{0,0}t^2_{1,1}-t^2_{1,0}t^2_{0,1}=\big(t^1_{1,0}\big)^2.
\end{gather*}

Recall its  Lax pair
\begin{gather*}
P_{1,0}=AP,\qquad P_{0,1}=BP,
\end{gather*}
where $P=(\psi^0,\psi^1,g^0,g^1)^T$ and
\begin{gather*}
A=\left(
\begin{array}{@{}c@{\,\,}c@{\,\,}c@{\,\,}c@{}}
\frac{t^1_{1,0}}{t^1_{0,0}}&-1&0&0\\
0&\frac{t^{2}_{1,0}t^1_{0,0}}{t^{2}_{0,0}t^1_{1,0}}&\frac{t^1_{-1,0}}{t^1_{0,0}}&
-\frac{t^1_{1,0}t^2_{-1,0}}{t^1_{0,0}t^2_{0,0}}\\
0&0&\frac{t^1_{-1,0}}{t^1_{0,0}}&0\\
0&0&-\frac{t^1_{-1,0}}{t^1_{0,0}}&\frac{t^1_{1,0}t^2_{-1,0}}{t^1_{0,0}t^2_{0,0}}
\end{array}\right),\qquad
B=\left(
\begin{array}{@{}c@{\,\,}c@{\,\,}c@{\,\,}c@{}}
1&0&0&0\\
\frac{t^2_{0,1}}{t^1_{0,0}t^1_{0,1}}&1&0&0\\
0&\frac{t^1_{0,0}}{t^1_{-1,1}t^2_{0,0}}&\frac{t^1_{0,1}t^1_{-1,0}}{t^1_{0,0}t^1_{-1,1}}&
\frac{t^2_{-1,0}t^2_{0,1}}{t^1_{-1,1}t^1_{0,0}t^2_{0,0}}\\
0&\frac{(t^1_{0,0})^2}{t^2_{-1,1}t^2_{0,0}}&0&
\frac{t^2_{-1,0}t^2_{0,1}}{t^2_{0,0}t^2_{-1,1}}
\end{array}\right).
\end{gather*}

By an automorphism $X\rightarrow \Phi X\Phi^{-1}$ transform the matrices $A$ and $B$ to the triangular matrices
\begin{gather}\label{bar_A}
\bar{A}=\left(
\begin{array}{@{}c@{\,\,}c@{\,\,}c@{\,\,}c@{}}
\frac{t^1_{1,0}}{t^1_{0,0}}&-1&0&0\\
0&\frac{t^{2}_{1,0}t^1_{0,0}}{t^{2}_{0,0}t^1_{1,0}}&
-\frac{t^1_{1,0}t^2_{-1,0}}{t^1_{0,0}t^2_{0,0}}&\frac{t^1_{-1,0}}{t^1_{0,0}}\\
0&0&\frac{t^1_{1,0}t^2_{-1,0}}{t^1_{0,0}t^2_{0,0}}&-\frac{t^1_{-1,0}}{t^1_{0,0}}\\
0&0&0&\frac{t^1_{-1,0}}{t^1_{0,0}}
\end{array} \right) ,\qquad
\bar{B}=\left(
\begin{array}{@{}c@{\,\,}c@{\,\,}c@{\,\,}c@{}}
1&0&0&0\\
\frac{t^2_{0,1}}{t^1_{0,0}t^1_{0,1}}&1&0&0\\
0&\frac{(t^{1}_{0,0})^2}{t^2_{-1,1}t^2_{0,0}}&\frac{t^2_{-1,0}t^2_{0,1}}{t^2_{0,0}t^2_{-1,1}}&0\\
0&\frac{t^{1}_{0,0}}{t^1_{-1,1}t^2_{0,0}}&\frac{t^2_{-1,0}t^2_{0,1}}{t^1_{-1,1}t^{1}_{0,0}t^2_{0,0}}
&\frac{t^{1}_{-1,0}t^{1}_{0,1}}{t^1_{-1,1}t^1_{0,0}}
\end{array} \right) .\!\!\!
\end{gather}

By automorphism $X\rightarrow F^{-1}XF$ transform the matrices (\ref{bar_A}) to the triangular matrices
\begin{gather*}
\hat{A}=\left(
\begin{array}{@{}c@{\,\,}c@{\,\,}c@{\,\,}c@{}}
\frac{t^1_{1,0}}{t^1_{0,0}}&-1&0&0\\
0&\frac{t^{2}_{1,0}t^1_{0,0}}{t^{2}_{0,0}t^1_{1,0}}&-\frac{t^1_{1,0}}{t^1_{0,0}}&1\\
0&0&\frac{t^{1}_{1,0}t^2_{0,0}}{t^1_{0,0}t^2_{1,0}}&-\frac{t^2_{0,0}}{t^2_{1,0}}\\
0&0&0&\frac{t^1_{0,0}}{t^1_{1,0}}
\end{array}\right),\qquad
\hat{B}=\left(
\begin{array}{@{}c@{\,\,}c@{\,\,}c@{\,\,}c@{}}
1&0&0&0\\
\frac{t^2_{0,1}}{t^1_{0,0}t^1_{0,1}}&1&0&0\\
0&\frac{(t^1_{0,0})^2}{t^2_{0,0}t^2_{0,1}}&1&0\\
0&\frac{t^1_{0,0}}{t^1_{0,1}t^2_{0,0}}&
\frac{t^2_{0,1}}{t^1_{0,0}t^1_{0,1}}&1
\end{array}\right).
\end{gather*}

Evaluate the elements at the right upper corner for the following products $\hat{A}_{2,0}\hat{A}_{1,0}\hat{A}_{0,0}$ and $\hat{A}_{3,0}\hat{A}_{2,0}\hat{A}_{1,0}\hat{A}_{0,0}$ and f\/ind two independent $m$-integrals
\begin{gather*}
I_{(1)}=-\frac{t^1_{3}}{t^1_{2}}-\frac{t^1_{2}t^2_{0}}{t^1_{1}t^2_{1}}-
\frac{t^1_{1}t^2_{2}}{t^1_{2}t^2_{1}}-\frac{t^1_{0}}{t^1_{1}}, \\
I_{(2)}= -\frac{t^1_4}{t^1_2}-\frac{t^1_1t^2_2t^1_4}{t^1_2t^2_1t^1_3}-\frac{t^1_1t^2_3}{t^1_3t^2_1}
-\frac{t^1_2t^2_0t^1_4}{t^1_1t^2_1t^1_3}-\frac{(t^1_2)^2t^2_0t^2_3}{t^1_1t^2_1t^1_3t^2_2}
-\frac{t^1_3t^2_0}{t^1_1t^2_2}-\frac{t^1_0t^1_4}{t^1_1t^1_3}-\frac{t^1_0t^1_2t^2_3}{t^1_1t^2_2t^1_3}
-\frac{t^1_0t^1_3t^2_1}{t^1_1t^2_2t^1_2}-\frac{t^1_0}{t^1_2}.
\end{gather*}
Replace the integral $I_{(2)}$ by a more simple one
\[\tilde I_{(2)}=I_{(2)}+I_{(1)}D_nI_{(1)}=\frac{t^2_{0}}{t^2_{1}}+\frac{(t^1_{1})^2t^2_{2}}{(t^1_{2})^2t^2_{1}}+
2\frac{t^1_{3}t^1_{1}}{(t^1_{2})^2}+\frac{(t^1_{3})^2t^2_{1}}{(t^1_{2})^2t^2_{2}}+
\frac{t^2_{3}}{t^2_{2}}.
\]
Here the second index for the variables $t^1$, $t^2$ is omitted, since its values is zero for all considered variables. It can be proved by using  Theorem~\ref{independent_int} that integrals $I_{(1)}$ and $\widetilde I_{(2)}$ constitute a~complete set of integrals.
\end{Example}

\section{Periodic boundary conditions}\label{section7}

In this section we discuss brief\/ly the well known periodically closed reduction of the Hirota chain (see for more details and the references \cite{Zabr}).  Close the chain (\ref{eqHir})  by imposing the periodical boundary conditions
\begin{gather*}
t^{-1}=t^{N}, \qquad t^{N+1}=t^{0}.
\end{gather*}
Close the Lax pair (\ref{lax1}) by setting the conditions on the eigenfunctions
\begin{gather}\label{periodic_lax}
\psi^{-1}=\lambda\psi^{N},\qquad \psi^{N+1}=\frac{1}{\lambda}\psi^{0}.
\end{gather}
As a result we get a f\/inite system of the form
\begin{gather}
t^{0}_{0,0}t^{0}_{1,1}-t^{0}_{1,0}t^{0}_{0,1}=t^{N}_{1,0}t^{1}_{0,1},\qquad
t^j_{0,0}t^j_{1,1}-t^j_{1,0}t^j_{0,1}=t^{j-1}_{1,0}t^{j+1}_{0,1},\qquad 1\leq j \leq N-1,\label{chain_5}\\
t^{N}_{0,0}t^{N}_{1,1}-t^{N}_{1,0}t^{N}_{0,1}=t^{N-1}_{1,0}t^{0}_{0,1},\nonumber
\end{gather}
which is closely connected with the Cartan matrix $A^{(1)}_N$.

Boundary conditions (\ref{periodic_lax}) reduce the sequence of linear discrete equations (\ref{lax1}) to a Lax pair for the reduced system (\ref{chain_5}). Introduce the eigenvector $P=(\psi^{0},\psi^{1},\dots,\psi^{N})^T$ and $N+1\times N+1$ matrices
\begin{gather*}
U=\left(
\begin{array}{@{}c@{\,\,}c@{\,\,}c@{\,\,}c@{\,\,}c@{}}
\frac{t^0_{0,0}t^1_{1,0}}{t^0_{1,0}t^1_{0,0}}&-1&0&\dots&0\\
0&\frac{t^{2}_{1,0}t^1_{0,0}}{t^{2}_{0,0}t^1_{1,0}}&-1&\dots&0\\
0&0&\frac{t^{3}_{1,0}t^2_{0,0}}{t^{3}_{0,0}t^2_{1,0}}&\dots&0\\
&&&\dots\\
-\frac{1}{\lambda}&0&\dots&0&\frac{t^{N}_{0,0}t^0_{1,0}}{t^{N}_{1,0}t^0_{0,0}}
\end{array}\right),\qquad
V=\left(
\begin{array}{@{}c@{\,\,}c@{\,\,}c@{\,\,}c@{\,\,}c@{}}
1&0&\dots&0&\lambda\frac{t^1_{0,1}t^N_{0,0}}{t^0_{0,0}t^0_{0,1}}\\
\frac{t^2_{0,1}t^0_{0,0}}{t^1_{0,0}t^1_{0,1}}&1&\dots&0&0\\
0&\frac{t^3_{0,1}t^1_{0,0}}{t^2_{0,0}t^2_{0,1}}&\dots&0&0\\
&&\dots\\
0&0&\dots&\frac{t^{N-1}_{0,0}t^0_{0,1}}{t^{N}_{0,0}t^{N}_{0,1}}&1
\end{array}\right).
\end{gather*}
If the functions $t^0,t^1,\dots,t^N$ solve the system (\ref{chain_5}) then the overdetermined system of linear equations
\begin{gather}\label{LPeriod}
P_{1,0}=UP,\qquad P_{0,1}=VP
\end{gather}
is compatible (see~\cite{Zabr}).

\begin{Remark}\label{period} Note that system (\ref{chain_5}) dif\/fers from the system corresponding to the same al\-geb\-ra~$A_N^{(1)}$ but def\/ined by the formula~(\ref{dis-dis-system}).
\end{Remark}

For example the system (\ref{chain_5}) for $N=2$ is of the form
\begin{gather}
t^0_{0,0}t^0_{1,1}-t^0_{1,0}t^0_{0,1}=t^2_{1,0}t^1_{0,1},\qquad
t^1_{0,0}t^1_{1,1}-t^1_{1,0}t^1_{0,1}=t^0_{1,0}t^2_{0,1},\nonumber\\
t^2_{0,0}t^2_{1,1}-t^2_{1,0}t^2_{0,1}=t^1_{1,0}t^0_{0,1}. \label{chain_per_3x3}
\end{gather}
In terms of the variables $u^0=-\log t^0$, $u^1=-\log t^1$, $u^2=-\log t^2$ it looks like
\begin{gather}
\Delta\big(u^0\big)=e^{u^0_{1,0}+u^0_{0,1}-u^1_{0,1}-u^2_{1,0}},\qquad
\Delta\big(u^1\big)=e^{-u^0_{1,0}+u^1_{1,0}+u^1_{0,1}-u^2_{0,1}},\nonumber\\
\Delta\big(u^2\big)=e^{-u^0_{0,1}-u^1_{1,0}+u^2_{1,0}+u^2_{0,1}}, \label{chain_per_2x2}
\end{gather}
while formula (\ref{dis-dis-system}) gives the system
\begin{gather*}
\Delta\big(u^0\big)=e^{u^0_{1,0}+u^0_{0,1}-u^1_{1,0}-u^2_{1,0}},\qquad
\Delta\big(u^1\big)=e^{-u^0_{0,1}+u^1_{1,0}+u^1_{0,1}-u^2_{1,0}},\\
\Delta\big(u^2\big)=e^{-u^0_{0,1}-u^1_{0,1}+u^2_{1,0}+u^2_{0,1}}.
\end{gather*}
After permutations $n\leftrightarrow m$ we get
\begin{gather}
\Delta\big(u^0\big)=e^{u^0_{1,0}+u^0_{0,1}-u^1_{0,1}-u^2_{0,1}},\qquad
\Delta\big(u^1\big)=e^{-u^0_{1,0}+u^1_{1,0}+u^1_{0,1}-u^2_{0,1}},\nonumber\\
\Delta\big(u^2\big)=e^{-u^0_{1,0}-u^1_{1,0}+u^2_{1,0}+u^2_{0,1}}. \label{chain_map_2x2}
\end{gather}

Obviously, systems \eqref{chain_per_2x2} and \eqref{chain_map_2x2} are dif\/ferent.

Let us give also the Lax pair for the system (\ref{chain_per_3x3}). Let us consider vector $P=(\psi^{0},\psi^{1},\psi^{2})^T$. Introduce $3\times 3$ matrices
\[
U=
\begin{pmatrix}
\frac{t^0_{0,0}t^1_{1,0}}{t^0_{1,0}t^1_{0,0}}&-1&0\\
0&\frac{t^2_{1,0}t^1_{0,0}}{t^2_{0,0}t^1_{1,0}}&-1\\
-\frac{1}{\lambda}&0&\frac{t^2_{0,0}t^0_{1,0}}{t^2_{1,0}t^0_{0,0}}
\end{pmatrix},
\qquad
V=
\begin{pmatrix}
1&0&\lambda\frac{t^1_{0,1}t^2_{0,0}}{t^0_{0,0}t^0_{0,1}}\\
\frac{t^2_{0,1}t^0_{0,0}}{t^1_{0,0}t^1_{0,1}}&1&0\\
0&\frac{t^0_{0,1}t^1_{0,0}}{t^2_{0,0}t^2_{0,1}}&1
\end{pmatrix}.
\]
If the functions $t^0$, $t^1$, $t^2$ satisfy the system (\ref{chain_per_3x3}) then the equations
\begin{gather*}
P_{1,0}=UP,\qquad P_{0,1}=VP.
\end{gather*}
are compatible.

\begin{Remark} The Lax pair (\ref{LPeriod}) can be rewritten in terms of the Cartan--Weyl basis of the algebra $A^{(1)}_N$
\begin{gather*}
P_{1,0}=\left(-\Lambda+e^{U_{1,0}-U_{0,0}}\right)P,\qquad
P_{0,1}=\left(E+e^{U_{0,1}}\bar{\Lambda}e^{-U_{0,0}}\right)P,
\end{gather*}
where $\Lambda=\sum\limits^N_{i=0}f_i$, $\bar{\Lambda}=\sum\limits^N_{i=0}e_i$, $U=\sum\limits^N_{i=0}u^ih_i$, $e^{-u^i}=t^i$,
\begin{gather*}
[h_i,h_j]=0,\qquad [e_i,f_j]=\delta_{ij}h_i,\qquad
[h_i,e_j]=A_{ij}e_j,\qquad [h_i,f_j]=-A_{ij}f_j,
\end{gather*}
and $A_{ij}$ are elements of the Cartan matrix of the algebra $A^{(1)}_N$.
\end{Remark}

\section{Conclusions}\label{section8}

A map is suggested converting any $N\times N$ matrix $A$ to a f\/inite system of dif\/ference-dif\/ference equations of exponential type (see~(\ref{dis-dis-system})). A hypothesis is formulated claiming that if~$A$ coincides with the Cartan matrix of a f\/inite or af\/f\/ine Lie algebra then the corresponding system is integrable. The hypothesis is approved by numerous examples. The systems obtained are rather simple and elegant. They essentially dif\/fer from those studied earlier (see survey~\cite{KNS} and references therein). For instance, the system corresponding the algebra~$G_2$ given in~\cite{KNS} reads~as
\begin{gather*}
T^{(1)}_m(u-1)T^{(1)}_m(u+1)=T^{(1)}_{m-1}(u)T^{(1)}_{m+1}(u)+T^{(2)}_{3m}(u),\\
T^{(2)}_{3m}\left(u-\frac{1}{3}\right)T^{(2)}_{3m}\left(u+\frac{1}{3}\right)=T^{(2)}_{3m-1}(u)T^{(2)}_{3m+1}(u)+
T^{(1)}_{m}\left(u-\frac{2}{3}\right)T^{(1)}_{m}(u)T^{(1)}_{m}\left(u+\frac{2}{3}\right),\\
T^{(2)}_{3m+1}\left(u-\frac{1}{3}\right)T^{(2)}_{3m+1}\left(u+\frac{1}{3}\right)\\
\qquad{}=T^{(2)}_{3m}(u)T^{(2)}_{3m+2}(u)+
T^{(1)}_{m}\left(u-\frac{1}{3}\right)T^{(1)}_{m}\left(u+\frac{1}{3}\right)T^{(1)}_{m+1}(u),\\
T^{(2)}_{3m+2}\left(u-\frac{1}{3}\right)T^{(2)}_{3m+2}\left(u+\frac{1}{3}\right)\\
\qquad{}=T^{(2)}_{3m+1}(u)T^{(2)}_{3m+3}(u)+
T^{(1)}_{m}(u)T^{(1)}_{m+1}\left(u-\frac{1}{3}\right)T^{(1)}_{m+1}\left(u+\frac{1}{3}\right),
\end{gather*}
while our formula (\ref{dis-dis-system}) generates $G_2$ system which can be represented as follows
\begin{gather*}
t^1_{n,m}t^1_{n+1,m+1}-t^1_{n+1,m}t^1_{n,m+1}=t^2_{n,m+1},\qquad
t^2_{n,m}t^2_{n+1,m+1}-t^2_{n+1,m}t^2_{n,m+1}=\big(t^1_{n+1,m}\big)^3.
\end{gather*}

In a recent article by Kimura, Yamashita and Nakamura \cite{KYN} a new application of conserved quantities of discrete-time integrable systems to numerical computations is suggested.  The systems studied in the present paper might have applications in such kind numerical methods.

\subsection*{Acknowledgments}

The authors are grateful to the  referees for their important contribution to improve the article. This work is partially supported by Russian Foundation for Basic Research (RFBR) grants 11-01-97005-r-povoljie-a, 12-01-31208-mol\_a and 10-01-00088-a and by Federal Task Program ``Scientif\/ic
and pedagogical staf\/f of innovative Russia for 2009--2013'' contract no. 2012-1.5-12-000-1003-011.

\pdfbookmark[1]{References}{ref}
\LastPageEnding

\end{document}